\documentclass[12pt,notoc]{JHEP3}
\usepackage{amsmath,amssymb,euscript,array,mathrsfs}
\usepackage{epsfig}
\newcommand{\beq}{\begin{equation}}
\newcommand{\eeq}{\end{equation}}
\newcommand{\bea}{\begin{eqnarray}}
\newcommand{\eea}{\end{eqnarray}}
\newcommand{\beqn}{\begin{eqnarray}}
\newcommand{\eeqn}{\end{eqnarray}}
\newcommand{\beas}{\begin{eqnarray*}}
\newcommand{\eeas}{\end{eqnarray*}}

\newcommand{\bquo}{\begin{quote}}
\newcommand{\enqu}{\end{quote}}

\newcommand{\startappendix}{
\setcounter{section}{0}
\renewcommand{\thesection}{\Alph{section}}}
\newcommand{\Appendix}[1]{
\refstepcounter{section}
\begin{flushleft}
{\large\bf Appendix \thesection: #1}
\end{flushleft}}
\def\Tr{ \hbox{\rm Tr}}

\def\stroke{\vrule height8pt width0.4pt depth-0.1pt}
\def\topfleck{\vrule height8pt width0.5pt depth-5.9pt}
\def\botfleck{\vrule height2pt width0.5pt depth0.1pt}
\def\Zmath{\vcenter{\hbox{\numbers\rlap{\rlap{Z}\kern 0.8pt\topfleck}\kern
2.2pt\rlap Z\kern 6pt\botfleck\kern 1pt}}}
\def\Qmath{\vcenter{\hbox{\upright\rlap{\rlap{Q}\kern
3.8pt\stroke}\phantom{Q}}}}
\def\Nmath{\vcenter{\hbox{\upright\rlap{I}\kern 1.7pt N}}}
\def\Cmath{\vcenter{\hbox{\upright\rlap{\rlap{C}\kern
3.8pt\stroke}\phantom{C}}}}
\def\Rmath{\vcenter{\hbox{\upright\rlap{I}\kern 1.7pt R}}}
\def\Z{\ifmmode\Zmath\else$\Zmath$\fi}
\def\Q{\ifmmode\Qmath\else$\Qmath$\fi}
\def\N{\ifmmode\Nmath\else$\Nmath$\fi}
\def\C{\ifmmode\Cmath\else$\Cmath$\fi}
\def\R{\ifmmode\Rmath\else$\Rmath$\fi}
\def\Tr{{\rm Tr}}

\def\2{{1\over 2}}

\def\4N{${\cal N}=4$}

\def\beq{\begin{equation}}
\def\eeq{\end{equation}}
\def\ba{\beq\new\begin{array}{c}}
\def\ea{\end{array}\eeq}

\title{Non-Abelian Vortices at Weak and Strong Coupling
  in  Mass Deformed ABJM Theory}
\author{Roberto Auzzi and S. Prem Kumar\\\\
{\it 
Department of Physics, \\
Swansea University, \\ 
Singleton Park, Swansea\\
SA2 8PP, U.K.  
}\\
E-mail: \email{r.auzzi}, \email{s.p.kumar@swansea.ac.uk}
} 
\abstract{We find half-BPS vortex solitons, at both weak and strong
  coupling, in the ${\cal N}=6$ supersymmetric mass deformation of
  ABJM theory with $U(N)\times U(N)$ gauge symmetry and Chern-Simons
  level $k$. The strong coupling gravity dual is obtained 
by performing a ${\mathbb Z}_k$ quotient of the ${\cal N}=8$
  supersymmetric eleven dimensional supergravity background of
  Lin, Lunin and Maldacena corresponding 
  to the mass deformed M2-brane theory.  At weak coupling, 
 the BPS vortices preserving  six supersymmetries are found in the
  Higgs vacuum of the theory where the gauge symmetry
  is broken to $U(1) \times U(1)$. The classical vortex solitons 
break a colour-flavour locked global 
  symmetry resulting in non-Abelian internal orientational
  moduli and a ${\bf CP}^1$ moduli space of solutions. 
At strong coupling and large $k$,
  upon reduction to type IIA strings, the vortex moduli space and its action
  are computed by a  probe D0-brane in the dual geometry. The mass of
  the D0-brane 
  matches the classical vortex mass. However, the gravity picture exhibits a six
  dimensional moduli space of solutions, a section of which can be
  identified as the ${\bf CP}^1$ we find classically, along with a
  Dirac monopole connection of strength $k$.
It is likely that the extra four dimensions in the moduli space are 
an artifact of the strong coupling limit and of the supergravity approximation.} 

\begin{document}
\section{Introduction}

The study of vortices with non-Abelian, internal
orientational degrees of freedom, has revealed beautiful connections
between their moduli space dynamics and features of the gauge theory
they live in
\cite{hananytong1,roberto1,shifmanyung,hananytong2}. Typically, such
classical solutions occur when the non-Abelian gauge symmetry is
spontaneously broken, and crucially, there exists a
``colour-flavour locked'' global symmetry in the vacuum. A vortex
solution breaking this colour-flavour symmetry then gives rise to a
continuous family of classical solutions which proves to be useful in
extracting vortex dynamics in the moduli space approximation. In this
paper we will investigate Chern-Simons vortex solitons in 2+1
dimensions, carrying non-Abelian internal orientational zero modes.
The theory we consider is a mass deformation of the ${\cal N}=6$
supersymmetric ABJM theory \cite{abjm} preserving all supersymmetries
\cite{massivo}.

Vortex solitons in Abelian and non-Abelian Chern-Simons theories have
been widely studied  
in both relativistic and non-relativistic settings
\cite{devega,csvortexother,csvortex,jackiwpi, kmlee,cssusy}. Detailed reviews
of these can be found in \cite{dunnerev,csvortexrev}. 
More recently, the moduli
space dynamics of (supersymmetric) non-Abelian 
Chern-Simons vortex solitons with
 internal collective coordinates, was analyzed in
\cite{schapo, ct, collie,gudnason}. It was already noted in \cite{kim} that the
Chern-Simons action induces terms which are first order in
time derivatives in the moduli space effective description of the
vortex. Specifically, the authors of \cite{ct, collie} demonstrated
that the effect 
of the Chern-Simons coupling on the moduli space quantum mechanics of
SUSY non-Abelian vortices, is to induce a coupling to a magnetic field
$\cal F$ which could then be given a geometric interpretation in terms
of the first Chern character of an index bundle over the moduli space.

One of our main motivations is to study the
semiclassical, solitonic objects arising in the context of the
recently discovered 
${\cal N}=6$ superconformal ABJM theory \cite{abjm} in 2+1
dimensions. The theory has a $U(N)\times U(N)$ gauge symmetry with
matter in the bifundamental representation and a level $(k,-k)$ Chern-Simons
action for the gauge fields. It describes the world-volume dynamics
of multiple M2-branes moving in a ${\mathbb C}^4/{\mathbb Z}_k$ 
orbifold background in M-theory. The ABJM proposal followed the
seminal works of Bagger-Lambert \cite{bl} and Gustavsson
\cite{gustavsson} (BLG), which first proposed the ${\cal N}=8$
superconformal theory on multiple M2-branes probing flat space.

We will see that the ABJM theory, when deformed by a particular
supersymmetric mass term, admits finite energy, non-Abelian 
Chern-Simons vortex solitons. What
makes the situation particularly interesting is that the soliton
dynamics can now be explored in two different regimes of the field
theory: one in which  semiclassical analysis is valid and
another wherein the theory is strongly coupled and is described by a
dual gravitational background. The study of vortices in these two
regimes, including the construction of the classical solution and
obtaining the dual gravity description at strong coupling, will be the
subject of the paper. Using these two approaches we confirm the
general picture of \cite{ct, collie}, while also encountering certain
unresolved puzzles. 

Various nonperturbative objects have been found in
BLG and ABJM theories. Monopole instantons in ABJM theory
 have been studied in \cite{nonpert}.
Vortex solitons have been already studied in 
mass deformed BLG theory in \cite{vortex1,vortex2}.
The solution found in \cite{vortex1} has a topological
winding and a mass that is twice the one found in \cite{vortex2}.
Vortices in mass deformed ABJM have been studied in \cite{arai};
those solutions can be interpreted as higher winding solutions
with respect to the ones that we will study in this paper.
Vortices in the non-relativistic limit of ABJM have been
studied in \cite{nonrel}.

Both the BLG theory in 2+1 dimensions and the
 ABJM theory admit mass deformations breaking conformal
invariance, but preserving all of their supersymmetries 
\cite{bena, bena1, llm, hosomichi, massivo}. In particular, the maximally
supersymmetric mass deformation of the ABJM theory was obtained in 
\cite{massivo, hosomichi} and the analysis of its classical vacuum
structure revealed a discrete set of vacua \cite{massivo}\footnote{A
  potential discrepancy was also noted,  
  since the classical vacuum states of the mass deformed theory are
  more numerous than expected from the
  supergravity dual.}. This deformation breaks the
$SU(4)\times U(1)$  global symmetry of the ABJM theory to 
$SU(2)\times SU(2)\times U(1)_A\times U(1)_B$.

We focus our attention on one of these classical
vacua which we expect to be perturbatively accessible for large $k$
(and $N/k \ll 1$),
and we refer to this vacuum as the ``Higgs vacuum''. Here the
$U(N)\times U(N)$ gauge group is broken to $U(1) \times U(1)$. We find that this
vacuum admits classical vortex solutions carrying both electric and
magnetic charge. Importantly, the Higgs vacuum exhibits a global
$SU(2)\times SU(2)_{C+F}\times U(1)_B  \times U(1)_A$ symmetry, where the
second $SU(2)$ factor arises via
 a combination of (broken) flavour $SU(2)$ rotations
and global gauge transformations. This colour-flavour locked
transformation acts non-trivially on our vortex solution which breaks
$SU(2)_{C+F}$ to $U(1)_{C+F}$, resulting in a ${\bf CP}^1$ moduli
space of solutions. We are able to construct the classical solutions
for all $N$, and show that they have finite energy and that they are
BPS. We explicitly check that the solutions are invariant under six
supercharges and are $\frac{1}{2}$-BPS states with a mass given by
$k\mu$, where $\mu$ is the mass deformation parameter of the theory.

The topology of the vacuum manifold ${\cal M}$ in the Higgs vacuum is
non-trivial, $\pi_1({\cal M})={\mathbb Z}_N$, and the vortex solitons carry
a ${\mathbb Z}_N$ charge.  However, they are actually stabilized also
by a global $U(1)_B$ charge which is quantized to be 
a multiple of $k$ \cite{abjm}, and which is not
carried by the perturbative states in the theory. Thus an $N$-vortex state
cannot annihilate into the vacuum. Although the vortex solution is
straightforward to obtain, its low energy dynamics on the moduli
space appears technically challenging to derive from first
principles. On general grounds, at weak coupling, since the
solutions preserve six supersymmetries and the moduli space of
solutions we have found is an $S^2$, we expect the moduli space
dynamics to be governed by supersymmetric quantum mechanics on a
sphere. However this leaves unclear, 
the effect of the Chern-Simons terms on this quantum
mechanics.

To learn more about the soliton dynamics we turn to the other
parametric regime where the mass deformed ABJM theory is
tractable. This is the strongly coupled, large $N$ limit, namely $N\to
\infty$, with $N/k$ large. In this limit the
${\cal N}=6$ superconformal ABJM theory is dual to eleven dimensional
supergravity 
on $AdS_4\times S^7/{\mathbb Z}_k$ obtained by a particular quotient
of the $AdS_4\times S^7$ solution dual to the ${\cal N}=8$
superconformal theory. We deduce the gravity dual of the mass deformed
ABJM theory by performing a similar quotient on the 
background dual to the maximally supersymmetric mass deformation of
the ${\cal N}=8$ superconformal M2-brane theory.  The latter
background, preserving ${\cal N}=8$ SUSY, and $SO(4)\times SO(4)$
symmetry, was obtained in \cite{bena,llm}. 
In the fermion fluid language of Lin, Lunin and Maldacena
\cite{llm}, the vacua of the $SO(4)\times SO(4)$, ${\cal N}=8$ theory 
are in one to one correspondence with partitions of $N$ and are
represented by states of free fermions. The Higgs
vacuum is the trivial partition and is a highly excited particle state
in the fermion picture.
This can be interpreted as the geometry generated by a dielectric
M5-brane carrying $N$ units of M2-brane charge and wrapped on one of the
two $S^3$'s in the $SO(4)\times SO(4)$ invariant geometry.

The quotienting of the $SO(4)\times SO(4)$ background above, by the ${\mathbb
Z}_k$ action, yields the Higgs vacuum of the mass deformed ABJM
theory, preserving an $SU(2)\times SU(2)\times U(1)_A\times U(1)_B$
symmetry. For large $k$ (such that $N/k$ is fixed and large), we can
reduce the geometry to type IIA string theory. The type IIA geometry
asymptotes to $AdS_4\times {\bf CP}^3$ and contains two spheres $S^2$
and $\tilde S^2$, each associated to one of the two $SU(2)$ factors of the
isometry group. The Higgs vacuum corresponds to a dielectric D4-brane
wrapping $S^2$ \cite{Myers}. 
The presence of the dielectric D4-brane can also be
directly inferred from a fuzzy sphere interpretation of the classical
VEVs in the Higgs vacuum \cite{nastase}. 
The general picture bears a strong resemblance to
the  ${\cal N}=1^*$ theory \cite{Donagi,Polchinski,dorey}, although
the geometries in the present situation are completely non-singular.
Non-Abelian vortices in the ${\cal N}=1^*$ theory where studied in \cite{yung,us}.

The vortex soliton in the Higgs vacuum is a D0-brane probe in the
above geometry\footnote{In the D-brane picture, the dielectric
  D4-brane will have a $B$-field along its worldvolume $S^2$ directions. 
This allows a
  D0-brane to form a bound state with the D4, corresponding to a
  noncommutative U(1) instanton in 5 dimensions, and appear as a
  vortex in the non-compact 2+1 dimensions.}. 
Surprisingly, we find that the probe mass is minimized
along a {\em six} 
dimensional submanifold ${\cal P}$, 
%which may be viewed as a
%deformation of ${\bf CP}^3$, 
preserving the reduced set of
isometries. The value of the probe mass along the moduli space ${\cal
  P}$ matches the value $\mu k$ deduced classically. The probe moduli
space ${\cal P}$ can be viewed as $S^2\times \tilde S^2\times S^1$
fibred along a segment ${\cal C}$, where the $S^1$ is also non-trivially fibred over the two $S^2$'s.
The topology  of a section at a generic point of the segment is  $S^2 \times S^3$.
 At the two tips of ${\cal C}$,  the three-sphere shrinks to zero 
\footnote{We thank D. Tong for drawing our attention to this.} and 
 the section is given by a two-sphere.
We identify the tip where
 the  $S^3$ obtained by fibering
$S^1$ over  $\tilde{S}^2$ 
 shrinks as the moduli space of vortex solutions we saw at weak
coupling. The probe dynamics in this section of ${\cal P}$ is that of
a particle on $S^2$ of radius $\sqrt {k/2}$ coupled to a Dirac
monopole connection of strength $k$. The radius of the sphere also
matches that of the fuzzy sphere from the classical analysis of the
Higgs vacuum. The effect of the Chern-Simons interactions on the
moduli space of the soliton, is to induce a Dirac monopole
connection. This picture is in agreement with the general results of 
\cite{ct, collie}.  However, the full six
dimensional moduli space at strong coupling presents a puzzle, and 
does not appear to have a
simple interpretation in terms of the soliton solutions we found at
weak coupling.

The paper is organized as follows. In Section 2, we review the
essential features of the ABJM theory and its mass deformation, their
symmetries, vacuum structure and equations of motion. Importantly, we
describe the origin of the colour-flavour locked symmetry in the Higgs
vacuum. In Section 3, we present our ansatz for the vortex soliton
solutions for general $N$ and discuss their stability and verify
explictly that they are left invariant by six supersymmetries. In
Section 4, we turn to the gravity dual of the mass deformed ABJM
theory. We first review the basic features of the $SO(4)\times SO(4)$
symmetric solution of \cite{bena, llm} and then explain the
quotienting procedure that yields the mass deformed ABJM theory. The
D0-brane probe dynamics and its moduli space are then deduced
straightforwardly.  We summarize our results and conclusions in
Section 5. In Appendix A  the vortex fermionic zero modes are discussed
for $N=2$.

{\bf Note Added:} While this paper was being completed, a closely
related preprint  arXiv:0905.1759 [hep-th] \cite{kkkn} appeared, 
which overlaps with our classical field theory analysis of the vortex
solitons. 

\section{Mass deformed ABJM theory}

The bosonic part of the Lagrangian of the ABJM theory \cite{abjm} is
given by a $U(N) \times U(N)$ Chern-Simons theory, coupled to
bifundamental matter with a scalar potential. The Chern-Simons levels
associated to the two gauge groups are $+k$ and $-k$ respectively.
In ${\cal N}=2$ superspace notation 
the ABJM superpotential for the bifundamental matter fields reads
\beq W=\frac{2 \pi}{k} \Tr (Q^1 (R^1)^\dagger Q^2 (R^2)^\dagger
-Q^1 (R^2)^\dagger Q^2 (R^1)^\dagger) \, .\eeq
where $Q^\alpha$ and $R^\alpha$ 
transform in the $({\bf N, \bar{N}})$ representation of the gauge
group. The global $SU(4)$ R-symmetry becomes explicit upon introducing
the fields
\beq C^I=(Q^1,Q^2,R^1,R^2) \,, 
\quad(I=1,\ldots 4),\eeq
and the bosonic part of the ABJM Lagrangian becomes
\begin{eqnarray}
{\cal L}_{\rm bosonic}=&& \frac{k}{4 \pi}  \epsilon^{\mu \nu \lambda} 
 \Tr \left( A_\mu \partial_\nu A_\lambda +\frac{2 i}{3} A_\mu A_\nu A_\lambda -
\hat{A}_\mu \partial_\nu \hat{A}_\lambda -\frac{2 i}{3} \hat{A}_\mu
\hat{A}_\nu \hat{A}_\lambda  
\right)\\\nonumber
&&-\Tr |D^\mu C^I|^2 +\frac{4 \pi^2}{3 k^2} \Tr \left( C^I C^\dagger_I
  C^J C^\dagger_J C^K C^\dagger_K +    
 C^I C^\dagger_J C^J C^\dagger_K C^K C^\dagger_I + \right. \\\nonumber
&&\left. +4   C^I C^\dagger_J C^K C^\dagger_I C^J C^\dagger_K -
6  C^I C^\dagger_J C^J C^\dagger_I C^K C^\dagger_K   
\right) \, ,
\end{eqnarray}
which is manifestly invariant under the $SU(4)$ R-symmetry
associated to  ${\cal N}=6$ supersymmetry.
The covariant derivatives on the bifundamental fields are defined as
\beq D^\mu C^I=\partial^\mu C^I +i A^\mu C^I- i C^I \hat{A}^\mu \, .\eeq
The fermionic part of the Lagrangian is,
\begin{eqnarray}
{\cal L}_{\rm fermionic}=&& -i \, \Tr (\psi^\dagger)^I \gamma^\mu D_\mu \psi_I
 +\frac{2 \pi i}{k} \, \Tr \left( C^\dagger_I C^I (\psi^\dagger)^J \psi_J - 
(\psi^\dagger)^J C^I C^\dagger_I \psi_J 
\right.
\\\nonumber
&& \left. -2 C^\dagger_I C^J (\psi^\dagger)^I \psi_J
+2 (\psi^\dagger)^J C^I C^\dagger_J \psi_I + \epsilon^{IJKL} C^\dagger_I \psi_J C^\dagger_K \psi_L  \right.
 \\\nonumber
&&\left. -  \epsilon_{IJKL} C^I (\psi^\dagger)^J C^K (\psi^\dagger)^L
 \right)  \, .
\end{eqnarray}
The conventions for $\gamma$-matrices are as in \cite{Benna:2008zy}: 
\beq \gamma^\mu = (i \sigma_2, \sigma_1, \sigma_3) \, .\eeq
To raise and lower spinor indices the $\epsilon^{\alpha \beta}$ symbol is used, with
 $\epsilon^{12}=-\epsilon_{12}=1$.
The charge conjugation on spinors is given by $\psi^c=\psi^*$. 
The metric choice is $g_{\mu \nu}= {\rm diag (-1,+1,+1)}.$

In \cite{massivo}, a mass deformation of the ABJM theory was found
which  preserves all the supersymmetries and 
breaks the $SU(4)_R\times U(1)$ global symmetry down 
to $SU(2) \times SU(2) \times U(1)_A \times U(1)_B \times {\mathbb Z}_2$. 
The ${\mathbb Z}_2$ action swaps the matter fields $Q^\alpha$ and
$R^{\alpha}$, while the $SU(2)$ factors act individually on the
doublets $\{Q^\alpha\}$ and $\{R^\alpha\}$ respectively.
The $U(1)_A$ symmetry rotates $Q^{\alpha}$ with a phase $+1$
and $R^{\alpha}$ with a phase $-1$.
This perturbation can be written as a superpotential in the $\mathcal{N}=1$
superfield formalism discussed in \cite{hosomichi}.
The R-symmetry group  is  $SU(2) \times SU(2) \times U(1)_A$.
This mass deformed theory is an example of three dimensional
supersymmetric theory with the so called "non-central" term in
the supersymmetry algebra \cite{nahm,linma} ; this means that the 
anticommutator of the supercharges closes not only in a combination of
momentum generators and central charges, but also in 
generators of the R-symmetry.
The expression in component fields is:
\begin{eqnarray}
\delta \mathcal{L}_{\rm mass} = && 
\mu^2 \Tr (Q^\alpha Q^\dagger_\alpha + R^\alpha R^\dagger_\alpha)
+\mu \, \frac{8 \pi }{k } \Tr (Q^\alpha Q^\dagger_{[\alpha} Q^\beta Q^\dagger_{\beta]}
-R^\alpha R^\dagger_{[\alpha} R^\beta R^\dagger_{\beta]}) -
\\\nonumber
&& - i \, \mu \, \Tr (   \xi^\dagger_1 \xi_1+ \xi^\dagger_2 \xi_2 -
 \chi^\dagger_{1} \chi_{1} - \chi^\dagger_{2} \chi_{2}  )  \, ,
\end{eqnarray}
where $\psi_I=(\xi_1,\xi_1,\chi_{1},\chi_{2})$.
The scalar potential of the mass deformed theory can
be written in a compact way as
\beq V =\Tr( |M^\alpha|^2+|N^\alpha|^2 )\, ,\eeq
where
\[  M^\alpha = \mu Q^\alpha +\frac{2 \pi}{k}
(2 Q^{[\alpha}  Q^\dagger_\beta Q^{\beta]}
+R^\beta R^\dagger_\beta Q^\alpha -Q^\alpha R^\dagger_\beta R^\beta+
2 Q^\beta R^\dagger_\beta R^\alpha - 2 R^\alpha R^\dagger_\beta Q^\beta) \, ,
\]
\[  N^\alpha = -\mu R^\alpha +\frac{2 \pi}{k}
(2 R^{[\alpha}  R^\dagger_\beta R^{\beta]}
+Q^\beta Q^\dagger_\beta R^\alpha -R^\alpha Q^\dagger_\beta Q^\beta+
2 R^\beta Q^\dagger_\beta Q^\alpha - 2 Q^\alpha Q^\dagger_\beta R^\beta) \, .
\]
It is also possible to consider the theory with gauge group
$SU(N) \times SU(N)$. In this case for $N=2$ 
we recover the Bagger-Lambert theory \cite{bl}. The $U(1)_B$ global
symmetry of the  $SU(N) \times SU(N)$ theory, is given
by the baryon number (under which $(Q^\alpha,R^\alpha)$ 
have charge $+1$). 

In the $U(N) \times U(N)$ gauge theory the naive baryon number symmetry is gauged
by a gauge field $A_b$ corresponding to the off-diagonal combination of the 
two Abelian factors in $U(N)\times U(N)$. The remaining Abelian symmetry 
$U(1)_{\tilde{b}}$ acts trivially on  all the matter fields and
couples to the theory through 
the Abelian Chern-Simons interaction ${\cal S}_{CS} = \frac{k}{4\pi} 
A_b \wedge F_{\tilde{b}}$. Hence there is another $U(1)_B$ global symmetry
generated by  the current $*F_{\tilde b}$ which is related by the 
the equation of motion for $A_b$ to the $U(1)_b$ 
current,
\beq J_\mu=\frac{k}{4 \pi}
\epsilon_{\mu \nu \rho} F^{\nu \rho}_{\tilde{b}} \, . \label{u1sym}\eeq
The flux quantization condition on $F_{\tilde{b}}$
implies that the $U(1)_B$ charges are quantized as integer multiples of
$k$. In the ABJM theory, the chiral
primary operators made from elementary fields, of the form ${\rm
  Tr}((C_I C_J^\dagger)^\ell)$, do not carry this $U(1)_B$ charge. Gauge
invariant operators carrying the quantized baryon number correspond to
combinations of the form $C^{nk}$ along with 
't Hooft operators. The presence of this global charge under which
elementary states are uncharged will be important for the stability of
the vortex solitons we find in the mass deformed theory below.

\subsection{Vacua and symmetries}
After mass deformation, the ABJM theory has several isolated
classical vacua preserving different amounts of gauge symmetry. 
These were obtained in \cite{massivo}. As in the case of the 
${\cal N}=1^*$ theory in $3+1$ dimensions \cite{Donagi, Polchinski},
classical vacua may be enumerated by finding block diagonal solutions
to the F-term vacuum conditions. In this case, for the scalar
potential to vanish we must have 
\beq
M^\alpha= N^\alpha=0.
\eeq
These equations have simple solutions if we assume that either
$R^\alpha=0$ or $Q^\alpha=0$. In the following we will concentrate on
some configurations with $R^\alpha=0$. This choice breaks the discrete
${\mathbb Z}_2$ symmetry. The potential for such configurations is 
\beq V=\Tr \left|\mu Q^\alpha+\frac{2 \pi}{k }(Q^\alpha Q^\dagger_\beta Q^\beta-
Q^\beta Q^\dagger_\beta Q^\alpha) \right| ^2  \, .\eeq

We consider the following vacuum, which corresponds to an
$N\times N$ irreducible solution and we will call this the Higgs
vacuum,
\begin{eqnarray}
Q^1=\sqrt{\frac{k \mu}{2 \pi}} \left(\begin{array}{ccccc}
0  &  &  & & \\
  & 1 &  & & \\
  &  & \ddots  & & \\  
    &  &  & \sqrt{N-2}& \\ 
        &  &  & & \sqrt{N-1}\\ 
\end{array}\right), 
  Q^2=\sqrt{\frac{k \mu}{2 \pi}} \left(\begin{array}{ccccc}
0  &  &  & & \\
 \sqrt{N-1} & \ddots &  & & \\
  & \ddots & 0  & & \\  
    &  & \sqrt{2} & 0& \\ 
        &  &  & 1 & 0\\ 
\end{array}\right) \, .   
\label{higgsvevs}
\end{eqnarray}
In this vacuum the gauge symmetry is almost completely broken by the
VEV,
\beq
{
U(N)\times U(N) \rightarrow U(1)_{\tilde{b}} \times U(1).
}
\eeq
It is a trivial fact that the $U(1)_{\tilde{b}}$ factor cannot be broken,
because it couples to the other fields of the theory
 just through Chern-Simons interactions.
If we label the two different gauge groups as $U(N)_L$ and $U(N)_R$,
the Higgs vacuum configuration breaks $U(1)_b$ and the $SU(N)_R\subset
U(N)_R$. The unbroken $U(1)$ gauge symmetry 
is a particular combination of the $U(1)_b$ with 
a diagonal generator of the $SU(N)_L$ gauge group which acts on
$Q^\alpha$ from the left. All other generators of $SU(N)_L$ are
broken. The unbroken generator is 
\beq K_L= {\rm Diag } (1,0,\ldots,0). \eeq
The global $SU(2)$ symmetry acting on the doublet $(Q^1,Q^2)$ is also
broken.

Similar to the case of the ${\cal N}=1^*$ theory, the above solutions
can be interpreted as fuzzy complex coordinates, which can be
decomposed into real (Hermitian) coordinates $X_p$  as in
\cite{Benna:2008zy},
\beq
{
Q^1= X^1+ i X^2, \quad Q^2= X^3+ i X^4.
}  
\eeq
The Higgs vacuum configuration implies that 
\beq
Q_\alpha^\dagger Q^\alpha = {\bf 1}\,(N-1) \frac{\mu
  k}{2\pi}
\label{fuzzyeq}
\eeq 
which formally resembles a fuzzy $S^3$ equation. However, due to the
fact that $Q^1$ is Hermitian implying that $X^2=0$, one suspects that
the configuration actually describes a fuzzy two sphere. This latter
picture has been confirmed in \cite{nastase}. Qualitatively the
situation is somewhat similar to the Higgs vacuum of the ${\cal N}=1^*$
theory characterized by such a fuzzy sphere configuration which breaks
both a global flavour symmetry and the gauge group. There, a
combination of the broken gauge and flavour generators can be shown to
generate a ``colour-flavour'' locked symmetry \cite{yung,us} which leaves
the VEVs invariant.

One expects therefore that the Higgs vacuum of the mass deformed 
ABJM theory should have an unbroken global 
symmetry which is a combination of the broken gauge transformations 
and the broken global $SU(2)$ symmetry that acts on
the doublet $(Q^1, Q^2)$. Indeed, we find such a colour-flavour locked
global symmetry of the vacuum.
  
For every $N$, there is a special combination of the broken global symmetry
and of the broken gauge symmetry which is left unbroken by the VEV.
Let us first denote the three generators of $SU(2)$ in an 
irreducible representation of dimension $m$, as
$J^a_m$ (with $a=1 \ldots 3$)

Now consider the following
$SU(2)$ global transformation , acting on the $Q^\alpha$:
\beq 
\left(\begin{array}{cccc}
Q^1\\
Q^2
\end{array}\right) \rightarrow U_F\,.\,\left(\begin{array}{cccc}
Q^1\\
Q^2
\end{array}\right)\qquad
U_F=\exp \left( i\alpha_a J^a_2 \right) \,.  \eeq
It can be checked that such a global transformation of the VEVs can be
undone by embedding the global rotation into (constant) 
$SU(N)_L \times SU(N)_R$ gauge transformations:
\beq 
Q^\alpha \rightarrow W_L \,Q^\alpha\,W_R^\dagger
\eeq
where
\begin{eqnarray}
&&W_L= \left(\begin{array}{c|cc}
1  & 0   \\ \hline  0 & \exp \left( i \alpha_1 J^1_{N-1}- i 
\alpha_2 J^2_{N-1}  - i\alpha_3 J^3_{N-1} \right) \\
\end{array}\right) \, ,\\\nonumber\\\nonumber
&&W_R= \exp \left( -i\alpha_1 J^1_{N} + i\alpha_2 J^2_{N}  +i\alpha_3
  J^3_{N} \right) \, . 
\end{eqnarray}
Note that it is only the broken gauge transformations which are
involved in the colour rotation.
We denote this unbroken ``colour-flavour'' locked symmetry as
$SU(2)_{C+F}$. Thus the Higgs vaccum of the mass deformed ABJM theory
has this symmetry and excitations around this vacuum should fall into
multiplets of the $SU(2)_{C+F}$ symmetry. A nice explanation of how
this embedding of global rotations in the gauge group is made possible, is
given in \cite{nastase}
\footnote{
As explained in \cite{nastase}, the set of matrices $J_\alpha^\beta = 
\frac{2\pi}{\mu k} Q_\alpha^\dagger Q^\beta$ are generators of
$U(2)$. If we further define $J_i = (\sigma^T_i)^\alpha_\beta \,
J_\alpha^\beta$, it is easily checked that these satisfy $SU(2)$
commutation relations. The $J_i$ transform as adjoints of the $U(N)_R$
gauge symmetry and provide an $N$-dimensional irreducible
representation of the $SU(2)$ algebra.
One may do the same with the matrices $\bar J_\alpha^\beta =
\frac{2\pi}{\mu k} Q^\beta Q_\alpha^\dagger$ and define $\bar J_i =
(\sigma_i)^\alpha_\beta\, \bar J _\alpha^\beta$. These furnish an $N-1$
dimensional irreducible representation and are adjoints under the
$U(N)_L$ gauge symmetry. The action of these generators of the gauge
symmetry on the bifundamentals $Q_\alpha$ precisely matches a global
$SU(2)$ rotation.
}. We will be interested in vortex solitons in
this vacuum and the existence of the colour-flavour locked symmetry has
interesting implications for the solitons.

\subsection{Equations of motion}

The classical picture for the vacuum states of the theory above and
the classical solutions we now look for, will only be valid in the
weakly coupled regime which in turn implies $k\gg 1$.
Since the ABJM theory has no Maxwell terms for the gauge fields, the
equations of motion for the gauge field yield Gauss law type constraints.
These are of the form
\begin{eqnarray}
\frac{k}{4 \pi} \epsilon^{\mu \nu \rho} F_{\nu \rho} =
i \left( (Q^\alpha) (D^\mu Q^\alpha )^\dagger-(D^\mu Q^\alpha )
  (Q^\alpha)^\dagger \right) \, ,\\\nonumber  
 \frac{k}{4 \pi} \epsilon^{\mu \nu \rho} \hat{F}_{\nu \rho} =  
 i \left( (D^\mu Q^\alpha )^\dagger (Q^\alpha) - (Q^\alpha)^\dagger
   (D^\mu Q^\alpha ) 
 \right) \, ,
\end{eqnarray}
where the field strength is defined as:
\beq F_{\mu \nu}=\partial_\mu A_\nu-\partial_\nu A_\mu + i [ A_\mu,A_\nu] \, .\eeq
Finally there are the  second order equations of motion for the scalar
fields $Q^\alpha$, 
with the ansatz $R^\alpha=0$,
\begin{eqnarray}
D_\mu D^\mu Q^1= &&\mu W^1+ \frac{2 \pi}{k}  \left( W^1 (Q^2)^\dagger
 Q^2 - Q^2 (Q^2)^\dagger W^1\right)  +
\\\nonumber 
 &&
 + \frac{4 \pi }{k} \mu \left( Q^1 (Q^2)^\dagger Q^2 -Q^2
 (Q^2)^\dagger Q^1 \right) + \\\nonumber 
 && +\frac{4 \pi^2}{k^2} \left( 
Q^1 (Q^1)^\dagger Q^1 (Q^2)^\dagger Q^2 + Q^1 (Q^2)^\dagger Q^2
 (Q^1)^\dagger Q^1+ 
Q^2 (Q^2)^\dagger Q^1 (Q^1)^\dagger Q^1 + \right. \\\nonumber
&&\left. + Q^1 (Q^1)^\dagger Q^2 (Q^2)^\dagger Q^1 - 2 Q^1
 (Q^2)^\dagger Q^1 (Q^1)^\dagger Q^2 - 
2 Q^2 (Q^1)^\dagger Q^1 (Q^2)^\dagger Q^1  \right) \, , 
\end{eqnarray}
where 
\beq W^1= \mu Q^1 +\frac{2 \pi}{k} \left(Q^1 (Q^2)^\dagger Q^2 -Q^2
  (Q^2)^\dagger Q^1 \right) \, , \eeq 
\[ W^2=\mu Q^2 +\frac{2 \pi}{k} \left(  Q^2 (Q^1)^\dagger Q^1- Q^1
  (Q^1)^\dagger Q^2 \right) \, . \] 
The equation of motion for $Q^2$ is identical to this and 
and can be obtained from the above by exchanging all $Q^1$'s with $Q^2$'s.

We will look for static, axially symmetric solutions to the above equations of
motion carrying charge under the $U(1)_B$ symmetry generated by $*F_{\tilde b}$. 
To this end we set the time derivatives of
$Q^\alpha$ and $A_{r,\varphi}$ to zero where $(r,\varphi)$ are polar
coordinates on the plane. In addition we choose the
gauge $A_r=0$.
We then get the following constraints
between $(A_0,\hat{A}_0)$ and $(F_{12},\hat{F}_{12})$ which are
particularly useful in solving for the scalar potentials, since they are
only algebraic conditions on the latter,
\begin{eqnarray} 
&&\frac{k}{2 \pi} F_{12} =   \left( Q^\alpha (Q^\alpha)^\dagger A^0
  +A^0 Q^\alpha (Q^\alpha)^\dagger 
 -2 Q^\alpha \hat{A}^0 (Q^\alpha)^\dagger \right)   \,
  ,\label{gausslaw} \\\nonumber  
&&\frac{k}{2 \pi} \hat{F}_{12} = \left( -  (Q^\alpha)^\dagger Q^\alpha
  \hat{A}^0 -\hat{A}^0 (Q^\alpha)^\dagger Q^\alpha  
 +2 (Q^\alpha)^\dagger   A^0 Q^\alpha  \right)  \, .
\end{eqnarray}
These relate the non-Abelian charge densities to the magnetic flux
carried by the configuration, in each gauge group factor. 
In our vortex ansatz we can use this constraint to fix 
the form of $A_0$ once that we have fixed an ansatz for $A_{\varphi}$
and for $Q^{\alpha}$. Note that this is a first order equation, but 
determines $A_0$ algebraically. 
The second set of Gauss constraints relate $F_{0r}$  to $A_{\varphi}$,
yielding a second order differential equation. In addition to this we
also need to ensure that the conditions $F_{0 \varphi}=0$ (and $A_r=0$)
emerging as a consequence of azimuthal symmetry are consistently satisfied.

Once we obtain explicit solutions to the Lagrange equations of motion,
we can compute the mass of the soliton using the following
expression for the energy density
\beq 
E=\int d^2 r (|D^0 Q^\alpha|^2 + |\vec{D} Q^\alpha|^2 + V(Q^\alpha)) \, .
\eeq

\section{Vortex in the Higgs vacuum}

Let us first of all discuss the topology of the vacuum manifold
\beq \mathcal{M}=\frac{SU(N)_L \times SU(N)_R \times U(1)_b}{U(1)_{\rm unbroken}}
=\frac{G}{H} \, .\eeq
There is a subtlety in the definition of $G$ that we now need to note. 
One combination of the centers of the $SU(N)_L$ and $SU(N)_R$ 
acts non-trivially on the
matter fields and this action can be undone by a ${\mathbb Z}_N$
rotation in $U(1)_b$. The other combination results in a 
${\mathbb Z}_N$ centersymmetry under which the matter fields are
uncharged.  Due to this reason, the fundamental group of $G$ is given by
$\pi_1(G)=\mathbb{Z} \oplus \mathbb{Z}_N$, where the $\mathbb{Z}_N$ factor corresponds
to non-contractible loops around which fields wind by a ${\mathbb Z}_N$ rotation 
generated by the diagonal combination of the centers of the $SU(N)_L$ and $SU(N)_R$
factors. We can then write the homotopy exact sequence:
\[ \ldots \rightarrow \pi_1\left( H \right) \rightarrow \pi_1\left(G\right)
\rightarrow
 \pi_1\left(G/H \right)\rightarrow \pi_0\left( H \right)
\rightarrow \ldots \]
\[ \ldots \rightarrow \mathbb{Z} \rightarrow \mathbb{Z} \oplus \mathbb{Z}_N 
\rightarrow
 \pi_1(\mathcal{M}) \rightarrow 0
\rightarrow \ldots \]
 From a straightforward application of the properties of the 
homotopy exact sequence,
it follows that
\beq \pi_1(\mathcal{M})=\mathbb{Z}_N \, .\eeq
The  vortex solitons are classified by a $\mathbb{Z}_N$
topological quantum number; if we take a configuration 
made of $N$ elementary vortices, they are not in principle any more 
topologically stable.

From the topological point of view, a configuration made by $N$
vortices actually  
corresponds to a trivial element of $\pi_1(\mathcal{M})$.
However, there is another quantum number that can
make the $N$-vortex configuration stable.  
As we have explained above, in the $U(N) \times U(N)$ gauge theory, 
the perturbative states of the theory are not charged under
the $U(1)_B$ global symmetry defined by the current in Eq.~(\ref{u1sym}).
The vortex solitons we find will be 
charged under this symmetry and for this reason
 protected from decaying to the perturbative states. 
 The $U(1)_B$ charge carried
 by these vortices is measured by the magnetic flux 
 associated to $*F_{\tilde b}$, carried by the soliton. These vortex
 solitons can also be thought of as states created by the 't Hooft
 monopole operators in the Higgs vacuum of the mass deformed theory.
See \cite{BKW} for a discussion of the corresponding
operators in conformal field theories.

% In contrast, in the $SU(N) \times SU(N)$ theory, both the perturbative
% states and the vortex soliton are charged under
% the global $U(1)$ baryonic symmetry.
% In this case a configuration of $N$ vortices is marginally stable,
% because the energy per unit of baryonic number 
% for the elementary charged particles and for the solitons
% is the same. This is very similar to the setting discussed in \cite{kmlee}.

In this section we will write an explicit ansatz for the elementary vortex
in the $U(N) \times U(N)$ theory. The vortex solutions in the
 $SU(N) \times SU(N)$ theory can be obtained by simply projecting out
the abelian part from the gauge fields $(A_\mu,\hat{A}_\mu)$. 

\subsection{Vortex Solution for $U(2) \times U(2)$ }

We begin with the simplest example with $N=2$.
In this case the solution that we find is very similar
to the one found in \cite{vortex2} 
in the mass-deformed Bagger-Lambert-Gustavsson theory,
which, for $N=2$  corresponds to ABJM theory with gauge group  $SU(2)
\times SU(2)$ \cite{vr}.  
The vortex ansatz should be
axially symmetric in two dimensions and the scalar field VEVs should
asymptote to the Higgs vacuum. We therefore take the ansatz,
 \beq Q^1= \sqrt{\frac{\mu k}{2 \pi}}
\left(\begin{array}{cc}
0  & 0 \\
 0  & 1 \\
\end{array}\right) \, ,  \qquad
 Q^2= \sqrt{\frac{\mu k}{2 \pi}}
\left(\begin{array}{cc}
0  & 0 \\
 e^{i \varphi} \psi(r)  & 0 \\
\end{array}\right)  \, .
\eeq
where the second scalar winds around origin. The ansatz breaks
completely, the $SU(2)_R \subset U(2)_R$ gauge symmetry which acts from
the right. A combination of the diagonal generator of $SU(2)_L$ with
$U(1)_b$ is however, preserved, while all fields are neutral under 
$U(1)_{\tilde b}$ . The vacuum manifold
\beq
{\cal M} = (SU(2)_L\times SU(2)_R\times U(1)_b)\big/U(1)
\eeq
has the fundamental homotopy group, $\pi_1({\cal M})={\mathbb Z}_2$. 
However, as we have already noted, the solutions with generic winding
numbers are stable due to the global $U(1)$ charge
associated to the symmetry generated by the current $*F_{\tilde b}$.
 
The spatial components of $A_\mu$ are  
\beq \hat{A}_i=A_{i}=
\frac{\epsilon_{ij} x_j}{r^2} \, (1-f(r)) \, 
 \frac{{\bf 1}_2  - \sigma_3 }{2}=  \frac{\epsilon_{ij} x_j}{r^2} \, (1-f(r)) \,  \left(\begin{array}{cc}
0  & 0 \\
0  & 1 \\
\end{array}\right) \, ,\eeq
from which  follows that the magnetic fluxes are
\beq F_{12}=\hat{F}_{12}=\frac{f'}{r} \frac{ {\bf 1}_2-\sigma_3}{2}=
\frac{f'}{r} \left(\begin{array}{cc}
0  & 0 \\
0  & 1 \\
\end{array}\right) \, . \eeq
Computing the charge associated to the $U(1)$ symmetry generated by the current
$*F_{\tilde b}$, we find
\beq
\int d^2x\;\frac{k}{2\pi} \epsilon_{0ij} F^{ij}_{\tilde b}= k,
\eeq
as expected for a state created by a 't Hooft operator.
The scalar gauge potentials  $(A_0,\hat{A}_0)$, are then determined by
the constraints  in Eq. (\ref{gausslaw}),  and are given by, 
\beq A_0=-\frac{f'}{\mu \, r}\frac{{\bf 1}_2-\sigma_3}{4}=
-\frac{f'}{2 \mu \, r} \left(\begin{array}{cc}
0  & 0 \\
0  & 1 \\
\end{array}\right)
 \, , \eeq
\[   \hat{A}_0=-\frac{f'}{\mu \, r}\frac{{\bf 1}_2+\sigma_3}{4}=
-\frac{f'}{2 \mu \, r} \left(\begin{array}{cc}
1  & 0 \\
0  & 0 \\
\end{array}\right)\,  .\]
Inserting the above ansatz into the equations of motion, 
we get the following equations for the vortex profile functions
$f(r)$ and $\psi(r)$,
\begin{eqnarray}
&&\psi '' +\frac{\psi '}{r} -\frac{f^2 \psi }{r^2} -2 \mu ^2
\psi(\psi^2-1)
= 0 \, , \label{nc2eqs}\\\nonumber
&& f'' -\frac{f'}{r}+ 4 f \mu ^2 \psi ^2 = 0 \, ,\\\nonumber 
&&\frac{\left(f'\right)^2}{4 r^2 \mu ^2}-\mu ^2 \left(\psi
  ^2-1\right)^2 =0 \, .
\end{eqnarray}
These equations are consistent and any two can be used to
derive the third. In fact they follow from first order BPS equations.
The BPS equations can be obtained by considering the energy functional 
\beq 
E=
\frac{k \mu^3}{2  \pi} \int 2 \pi r \, dr \, \left(  \frac{1}{4 \mu^4
  } \frac{(f')^2}{r^2} +   \frac{1}{\mu^2} \left( \frac{f^2 \psi^2}{ \, r^2}+
(\psi')^2   \right) +  (\psi^2-1)^2 \right)\, .
\eeq
Rearranging various terms we find the Bogomol'nyi completion,
\begin{eqnarray}
E= &&k  
\int 2 \pi r \, dr \left(  2 \left( \frac{f'}{4 r \sqrt{\mu  \pi}} -\frac{\mu^{3/2} (\psi^2-1)}{2 \sqrt{\pi}}\right)^2
+\left( \psi' - \frac{f \psi}{r}\right)^2 \frac{\mu}{2 \pi}  \right)
+\\\nonumber 
&& 
+ k \mu \, \int \, dr \, \partial_r \left(  f (\psi^2-1) \right)\, .
\end{eqnarray}
The first order equations obeyed by 
BPS solutions are equivalent to the three equations (\ref{nc2eqs})
which are then automatically satisfied.

We further note that, even though this is physically a Chern-Simons vortex,
the equations for the profiles are formally the same as the ones
for the BPS Abrikosov-Nielsen-Olesen vortex \cite{ANO}.
The magnetic field has a maximum at the origin, unlike
the conventional Chern-Simons vortex.
The BPS vortex mass is 
\beq 
T= k \mu \, . 
\eeq

Importantly, it is straightforward to check that the solutions above are
left invariant by the action of a $U(1)$ subgroup of the colour-flavour
locked symmetry $SU(2)_{C+F}$ in the Higgs vacuum. The unbroken
$U(1)_{C+F}$ is generated by the combined action of the diagonal generator
(proportional to $\sigma_3$) of the gauge $SU(2)_R$ and that of the 
$SU(2)$ R-symmetry which acts on the doublet $\{Q^\alpha\}$. Hence the
soliton is endowed with an internal moduli space
\beq
{\bf CP}^1\simeq SU(2)_{C+F}/U(1)_{C+F}.
\eeq
Acting on the vortex with the broken generators of $SU(2)_{C+F}$
generates new solutions and changes the orientation of the non-Abelian
flux within the $SU(2)_R$ gauge group factor. Note that this does not
change the global charge under $*F_{\tilde b}$.
\subsection{Vortex solution for $U(3) \times U(3)$}

We now exhibit the explicit ansatz and solution for the $N=3$ case.
This will provide some intuition for how to obtain the general 
solution. The ansatz for the bifundamental scalars approaching the
Higgs vacuum at infinity is
 \beq Q^1= \sqrt{\frac{\mu k}{2 \pi}}
\left(\begin{array}{ccc}
0  & 0 & 0\\
 0  & 1 & 0\\
 0  & 0 & \sqrt{2} \\
\end{array}\right) \, ,  \qquad
 Q^2= \sqrt{\frac{\mu k}{2 \pi}}
\left(\begin{array}{ccc}
0  & 0  &0\\
 \sqrt{2} \kappa(r)  & 0 & 0\\
0 & e^{i \varphi} \psi(r) & 0  \\
\end{array}\right) \,  ,
\eeq
where we have introduced one additional real profile function
$\kappa$ for the scalar that winds around the origin. We find that
$\kappa$ remains non-vanishing for all $r$. The spatial vector fields
have the form, 
\beq
\hat{A}_i=A_{i}=
\frac{\epsilon_{ij} x_j}{r^2} \, 
\left(\begin{array}{ccc}
0  & 0 & 0\\
 0  & - g(r) & 0\\
 0  & 0 & 1-f(r) \\
\end{array}\right)
 \, ,
\eeq
whilst the scalar gauge potentials are chosen to satisfy the Gauss law
constraints:
\beq
A_0=-\frac{1}{4 r \mu}\left(\begin{array}{ccc}
0  & 0 & 0\\
 0  & f'+2 g' & 0\\
 0  & 0 & f' \\
\end{array}\right) \, , \qquad 
\hat{A}_0=-\frac{1}{4 r \mu}\left(\begin{array}{ccc}
 f'+2 g' & 0 & 0\\
 0 & f' & 0 \\
 0  & 0 & 0\\
\end{array}\right) \, .
\eeq
The function $f$ approaches unity at $r=0$
and vanishes as $r \rightarrow \infty$. On the other hand, the profile function
$g(r)$ is zero both at $r = 0$ and at $r \rightarrow \infty$, and thus
does not influence the flux carried by the solution.

It is possible to obtain first order BPS equations for the ansatz by
expressing the vortex energy energy functional as a sum of
squares. The energy functional 
\begin{eqnarray}
E= \frac{k \mu}{2  \pi} \int 2 \pi r \, dr \, 
&&\left(  \frac{(f')^2+ 2 (g')^2}{8 r^2 \mu^2} + 
(\psi')^2+2 (\eta')^2+\frac{(f-g)^2 \psi^2 + 2 g^2 \kappa^2}{r^2}  \right.
\nonumber\\
&&\left. +  2  \mu^2 (\psi^2-1)^2 + \mu^2 (\psi^2 - 2 \kappa^2 +1)^2
 \right)\, .
\end{eqnarray}
It is fairly easy to the infer the Bogomol'nyi completion which
implies first order BPS equations,
\begin{eqnarray} 
E=\int 2 \pi r \, dr \, 
&&\left( \frac{k}{8 \pi \mu} \left( 2 \mu^2(1-2 \kappa^2+\psi^2
    )+\frac{g'}{r} \right)^2+ 
 \frac{k}{16 \pi \mu} \left( 4 \mu^2 (\psi^2-1)-\frac{f'}{f}
 \right)^2+ \right.\nonumber\\ 
&&\left. + \frac{k \mu}{2 \pi} \left( \psi' - \frac{(f-g) \psi}{r}\right)^2 +
\frac{k \mu}{ \pi} \left( \kappa' -\frac{g \kappa}{r} \right)^2
\right)  + \\\nonumber
&& +k \mu \int \, dr \, \partial_r \left(  f (\psi^2-1)+ g(2 \kappa^2-\psi^2-1) \right) \, .
\end{eqnarray}
The system of fist order BPS equations can be solved numerically;
the result is shown in Figure~1.
The new function $g$ does not influence the mass of the soliton
since it vanishes both at the origin and at infinity. Hence the 
vortex mass is again:
\beq T= k \mu \, . \eeq
It is straightforward to check that the gauge theory equations of motion 
are satisfied.

\begin{figure}[h]
\begin{center}
\epsfig{file=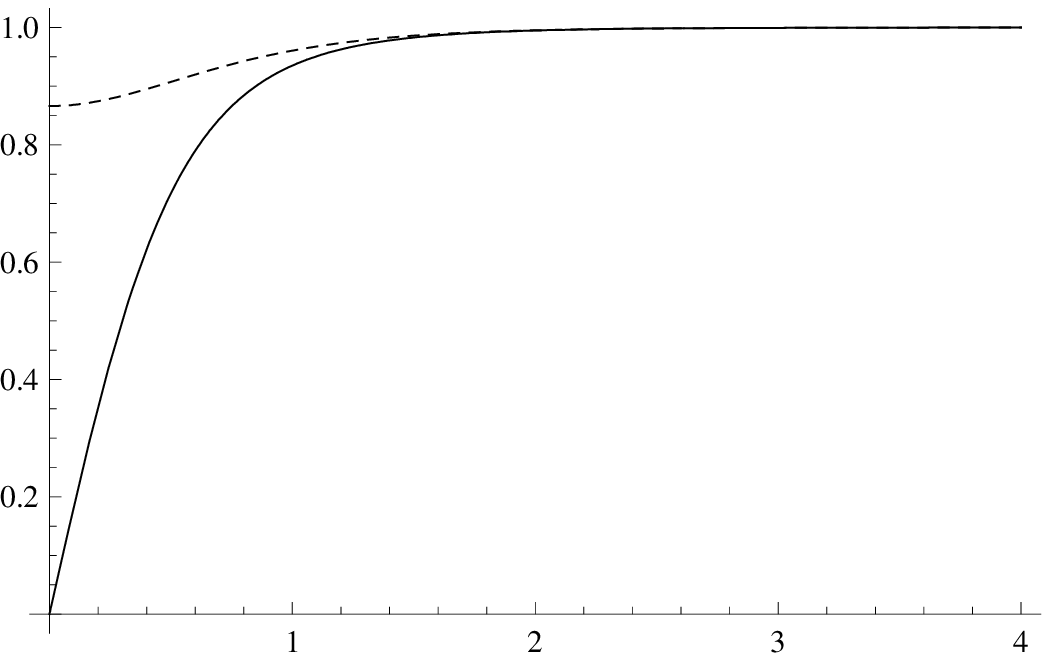, width =2.2in}\hspace{0.5 in}
\epsfig{file=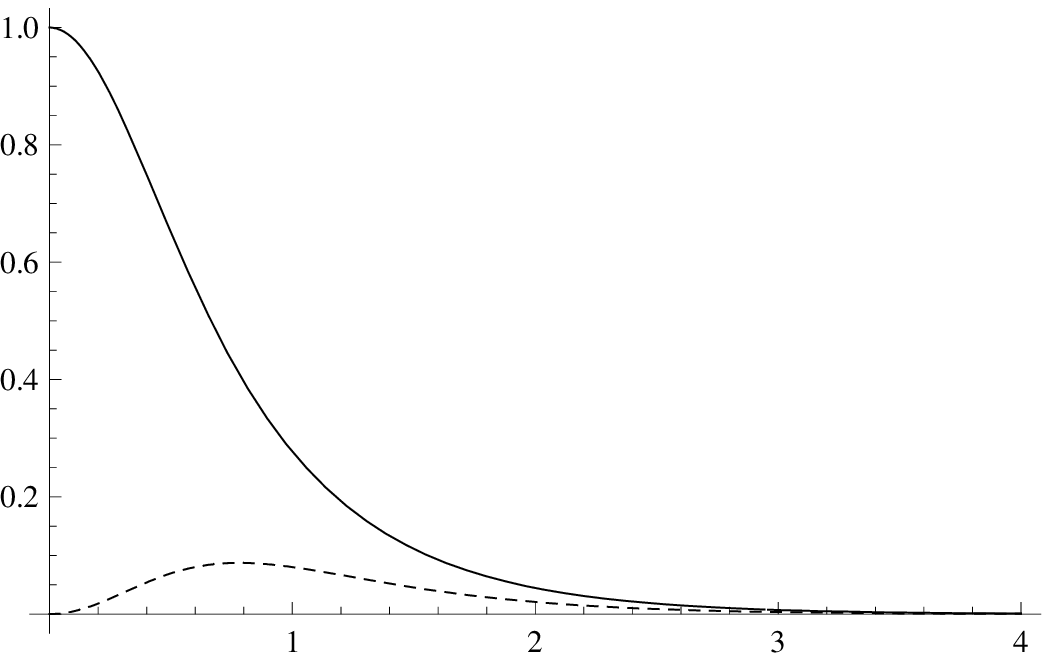, width=2.2in}
\end{center}
%\epsfxsize=6cm \centerline{\epsfbox{nc2.eps}}
%\noindent\small{\bf Figure 1:} 
\caption{The vortex profile for $N=3$. Left:
$\psi$ (solid), $\kappa$ (dashes). Right: $f$ (solid), $g$ (dashes).}
\label{nc3profiles}
\end{figure}

\subsection{Vortex solution for $U(N) \times U(N)$}

It is now straightforward to write the soliton ansatz for generic $N$.
The field $Q^1$ is taken to be constant and equal to its VEV in the
Higgs vacuum Eq. \eqref{higgsvevs}.
The non-zero entries of $Q^2$ are parameterized as:
\beq (Q^2)_{N,N-1} = \sqrt{\frac{\mu k}{2 \pi}} \, e^{i \varphi} \,
\psi(r) \, \qquad  
(Q^2)_{N-j,N-j-1} = \sqrt{\frac{\mu k}{2 \pi}} \,  \sqrt{j+1} \,
\kappa_j(r)  \, , 
\eeq
with $j=1,2,\ldots N-2$. The new radial profile functions will generically be
non-zero whn solved for. An additional set of 
$N-1$ functions is also necessary for
the gauge fields,
\beq A_i=\hat{A}_i=
\frac{\epsilon_{ij} x_j}{r^2} \,  {\rm Diag}  \, ( 0 , -g_{N-2}(r),
 \ldots, -g_{1}(r), 1-f(r)) \, . 
\eeq
Of these, only $f(r)$ influences the net magnetic flux, since the
$g_\ell$ vanish at the origin and at infinity. 
The time component of the gauge fields are given by:
\begin{eqnarray} 
&&A_0=\frac{-1}{2 \mu \, r} {\rm Diag} \,
 \left(0, \frac{f'}{N-1}+\sum_{j=1}^{N-2} \frac{g'_j}{N-1-j},
\frac{f'}{N-1}+\sum_{j=1}^{N-3} \frac{g'_j}{N-1-j}, \ldots ,
\frac{f'}{N-1} \right) \, , \nonumber\\\\\nonumber
&&\hat{A}_0=\frac{-1}{2 \mu \, r} {\rm Diag} \,
 \left( \frac{f'}{N-1}+\sum_{j=1}^{N-2} \frac{g'_j}{N-1-j},
\frac{f'}{N-1}+\sum_{j=1}^{N-3} \frac{g'_j}{N-1-j}, \ldots ,
 \frac{f'}{N-1},0 \right) \,. 
\end{eqnarray}
We have then to write $2(N-1)$ first order BPS equations for these
profile functions. From our solutions for $N=2$ and 3, we conclude
that the BPS solutions satisfy the equations,
\beq
  D_0 Q^1 - i W^1 = 0 \, , \qquad  \, D_1 Q^2 + i D_2 Q^2 =0 \, . \label{bp2}
\eeq
These equations lead to first order differential equations for the
profile functions. The following set of equations is also trivially
satisfied by our ansatz\"e,
\beq 
D_1 Q^1 = 0 \, , \qquad D_2 Q^1 = 0 \, , \qquad D_0 Q^2 = 0  \, ,
\qquad W^2 = 0 \, . \label{bp1} 
\eeq
Below, for completeness we list the first order equations of motion
for general $N$,
\begin{eqnarray}
&&\frac{f'}{r} + 2 \, (N-1) \, \mu^2 \,(1-\psi^2) =0 \, , \qquad
 \frac{g'_1}{r} + 2 \, (N-2) \, \mu^2 \, \,(1+\psi^2-2 \kappa_1^2) =0
 \, , \nonumber\\ 
&&\frac{g'_j}{r} + 2 \, (N-1-j) \, \mu^2 \, 
(1+ j \,\kappa_{j-1}^2-(j+1) \,  \kappa_j^2) =0 \, , \qquad
2 \leq j \leq N-2 \, .\\
&&\psi' - \frac{(f-g_1) \, \psi}{r} =0 \, , \qquad
 \kappa'_j - \frac{(g_j-g_{j+1}) \, \kappa_j}{r}=0 \qquad (1 \leq j \leq N-3) \, ,\\\nonumber
&&\kappa'_{N-2} - \frac{g_{N-2} \, \kappa_{N-2}  }{r}=0 \, .
\end{eqnarray}
It is straightforward to check that in general
\beq T= k \mu \, . \eeq

The global $SU(2)_{C+F}$ for the Higgs vacuum is broken by the vortex soliton
to $U(1)_{C+F}$. This
latter symmetry is generated by a combination of the diagonal
R-symmetry generator, along with the 
generator proportional to $(0,J_{N-1}^3)$ of the $SU(N)_L$ gauge group
and the generator $J^3_N$ of the $SU(N)_R$ gauge group
factor. Therefore, the vortex soliton for general $N$ also has a 
${\bf CP}^1$ moduli space for its internal orientational degrees of freedom.

% The vacuum for general $N$  break the gauge symmetry
% $U(N)_L\times U(N)_R$ to a $U(1)$ generated by the combination of a
% diagonal $SU(N)_L$ generator with the $U(1)_b$. 
% The fundamental group of the vacuum manifold is
% trivial and the solitons are non-topological. They are stable against
%  decay to perturbative states
%  due to their charge under $*F_{\tilde b}$. 

\subsection{BPS conditions and Supersymmetry Check }

In supersymmetric theories, the vortex first order equations are usually related
to some amount of preserved supersymmetry (see \cite{cssusy}
for a discussion in the case of the $U(1)$ Chern-Simons vortex). In
the case at hand we will see that our solutions preserve one half of
the supersymmetries of the full ${\cal N}=6$ supersymmetric 
mass deformed theory. To check the supersymmetry variations around the
soliton vortex solutions we will need the general SUSY variations of
the mass deformed ABJM theory.  The supersymmetry transformations of
the mass deformed theory differ very slightly from those of the
conformal theory. We follow the notation of \cite{terashima}
for the $\mathcal{N}=6$ SUSY transformations of the ABJM theory and
infer the effect of the mass deformation from the work of \cite{hosomichi}.
Let us check how many supersymmetries are preserved by the
vortex solution. 

In order to parameterize the $\mathcal{N}=6$ supersymmetries, 
let us introduce 6 Majorana real spinors $\epsilon_i$ \, $(i=1, \ldots ,6)$  
and use them to define $\omega_{AB}$, the spinor valued 
totally antisymmetric tensor of $SU(4)$,
\beq 
\omega_{AB} = \epsilon_i (\Gamma^i)_{AB} \, , \qquad 
\omega^{AB} = \epsilon_{i} ((\Gamma^i)^*)^{AB} \, ,
\qquad A,B=1,\ldots 4.
\eeq
Here $\Gamma^i$ are $SO(6)$ gamma matrices, 
represented as a set of anti-symmetric matrices \cite{terashima};
the conventions for the fermionic part of the lagrangian are the same as in \cite{Benna:2008zy}.
The explicit expression for $\omega_{AB}$ is
\beq \omega_{AB}=\epsilon_k \Gamma^k_{AB}= \left(\begin{array}{cccc}
0  & -\epsilon_6 -i \epsilon_5 &  \epsilon_3+i \epsilon_4 & -\epsilon_2 -i \epsilon_1\\
\epsilon_6 +i \epsilon_5  & 0 & \epsilon_2 -i \epsilon_1 & \epsilon_3-i \epsilon_4 \\
-\epsilon_3-i \epsilon_4  & -\epsilon_2 +i \epsilon_1 & 0 & \epsilon_6 -i \epsilon_5 \\
 \epsilon_2+i\epsilon_1  & - \epsilon_3+ i \epsilon_4 &  -\epsilon_6 +i \epsilon_5 & 0\\
\end{array}\right) \, .\eeq
With these conventions $\omega_{41}=\omega_{23}^*$, $\omega_{31}=\omega_{42}^*$
and $\omega_{43}=\omega_{12}^*$.
These provide a parametrization of the SUSY variations of the
$SU(4)$ R-symmetry invariant ABJM theory. 

The $\mathcal{N}=6$ SUSY transformations then read,
\begin{eqnarray}
&&\delta \psi_E= \nonumber\\\nonumber
&&\gamma^\mu \omega_{EF}  D_\mu C^F +
 \frac{2 \pi}{k} \left( -\omega_{EF} (C^G C^{\dagger}_G C^F - 
C^F C^\dagger_G C^G) + 2 \omega_{GH} C^G C^\dagger_E C^H \right) +
\nonumber\\
&&+ \mu \left( M_E^{\,\,\,F}\omega_{FG} \,C^G \right) \,
 ,\label{susytrasf}\\\nonumber\\\nonumber
&& M_{E}^{\,\,\,F} = {\rm Diag}(-1,-1,1,1),
\\\nonumber\\\nonumber
&&\delta A_\mu = -\frac{2 \pi}{k}  (C^E (\psi^{\dagger})^F \gamma_\mu
\omega_{EF} 
 +\omega^{EF} \gamma_\mu \psi_E C^\dagger_F  ) \, , \\\nonumber\\\nonumber
&&\delta \hat{A}_\mu = \frac{2 \pi}{k}  ((\psi^{\dagger})^E C^F
\gamma_\mu \omega_{EF} 
 +\omega^{EF} \gamma_\mu C^\dagger_E \psi_F ) \, , \qquad
 \delta C^E = i \omega^{EF} \psi_F  \, .
\end{eqnarray}
The matrix $M_{E}^{\,\,\,F}$ breaks the $SU(4)$ R-symmetry to
$SU(2)\times SU(2)$ and implements the mass deformation.
On our solutions which have $R^\alpha=0$,
the SUSY variations of the four fermionic fields become
\begin{eqnarray} 
&&\delta \psi_1=\gamma^\mu \omega_{12} D_\mu Q^2 - \omega_{12} W^2 \, ,\\
&&\delta \psi_2=  \gamma^\mu \omega_{21} D_\mu Q^1 - \omega_{21} W^1 \, ,\\
&&\delta \psi_3 =  \gamma^\mu \omega_{31} D_\mu Q^1 + \gamma^\mu \omega_{32} D_\mu Q^2 + \omega_{31} W^1 + \omega_{32} W^2 \, , \\
&&\delta \psi_4 =  \gamma^\mu \omega_{41} D_\mu Q^1 + \gamma^\mu
\omega_{42} D_\mu Q^2 + \omega_{41} W^1 + \omega_{42} W^2 \, . 
\end{eqnarray}

It is straightforward to check that, provided the equations (\ref{bp2},\ref{bp1}) 
are satisfied, the following 6 SUSY generators are unbroken:
\beq \omega_{12}= \left(\begin{array}{c} i \\ 1\end{array}\right)\alpha_1 \, , \qquad
\omega_{32}=\left(\begin{array}{c} i \\ 1\end{array}\right) \alpha_2 \, ,\qquad 
\omega_{42}=\left(\begin{array}{c} i \\ 1\end{array}\right) \alpha_3 \, ,  \label{nonrotte}
\eeq
where $\alpha_{1,2,3}$ are three complex grassmann numbers.
The vortex soliton is a $1/2$ BPS object preserving six supercharges.

It is important to stress that the ones in Eq.~(\ref{nonrotte})
are the unbroken supercharges for a vortex oriented in a specific direction in
the $SU(2)_{C+F}$ space. What we are calling $SU(2)_F$ is an R-symmetry
of the theory and is acting in a non-trivial way on the paremeters
$\omega_{AB}$, rotating the indices $A,B=1,2$ as an $SU(2)$ doublet
and acting trivially on $A,B=3,4$. As a consequence, if we rotate
the vortex in the $SU(2)_{C+F}$ space, we are changing the set of the supercharges
that are left unbroken by the vortex. 

\subsection{Comments on the vortex effective theory }
\label{caporetto}

In this section we have found a classical vortex solution
for arbitrary $k,N$ with minimal winding.
This object breaks spontaneosly the $SU(2)_{C+F}$
symmetry to $U(1)_{C+F}$. Due to this reason, acting with
the broken symmetry we can build an $S^2=SU(2)_{C+F}/U(1)_{C+F}$ 
moduli space of classical vortex solutions.
Our classical analysis above is valid when $k\gg 1$, for fixed $N$,
when the theory is weakly coupled\footnote{
 The $U(N) \times U(N)$ theory with
$k=1,2$ is supposed to have enhanced supersymmetry and
global symmetry; of course in this regime we cannot trust the semiclassical
approximation.  Also the case with $SU(2)\times SU(2)$ gauge symmetry, which corresponds to
the Bagger-Lambert theory, is different because there are extra global
symmetries and supersymmetries. 
}. In the large $N$ limit, the
semiclassical solutions can be trusted provided the 't Hooft coupling 
$\lambda = N/k \ll 1$.

In the next section we will see that the gravity dual,
which should be a good description of the physics at large $\lambda$,
suggests that this is not all of the story.
With this other approach a
 larger internal bosonic moduli space with dimension six is found for
the elementary vortex.
A possible interpretation of this result is 
 that our ansatz in field theory is not general enough
to accomodate the most general vortex solution.
In our calculation we keep always the scalars $R^\alpha=0$;
it is possible that a more general solution with non-zero $R^\alpha$ exists.  
Another possible interpretation is that the extra four dimensions 
of the moduli space found in the string theory dual are an artifact
of the strong coupling limit and of the supergravity approximation.
We believe that the former of the two options is unlikely,
as, after a fair amount of study, we were unable to 
arrive at a reasonable possible ansatz for a more general solution.
In order to solve the issue
 a detailed analysis of the bosonic zero modes of the solution
 along the lines of \cite{WeinbergA} 
should be performed. This analysis is not so straighforward,
because we have first to guess the form of the generalized
BPS equations, which are not completely obvious in this case.
 We leave this issue as a topic for further investigation.

Let us denote with $\mathcal{R}$ the vortex internal moduli space
($\mathcal{R}$ will include at least the $S^2$ moduli space that we have discussed in this section).
The vortex dynamics is then described by an effective quantum mechanics
with target space $\mathcal{R}$.
The effective one dimensional sigma model
involves not only second order term in the vortex velocities
 (the moduli space metric),
but also  first order term (which can be regarded as effective magnetic fields on the moduli space). 
These first order terms are a common feature of soliton dynamics
in Chern-Simons theories \cite{ct, kim}.
If $\mathcal{R}=S^2$,
we expect that the vortex dynamics is described by
the quantum mechanichs of a charged particle
on a 2-sphere in the background of the field of a 
magnetic monopole \cite{wuyang}. This basic picture appears to be
confirmed by our study of the dual gravity picture in the next section.

Since our solitons are  BPS objects and preserve 
some of the supersymmetries of the theory, we expect that 
the bosonic internal orientation moduli will be accompanied by 
fermionic super-orientational zero modes.
Monopole quantum mechanics with different
amounts of supersymmetries 
have been studied in \cite{monopolesusy} and  \cite{hllo}.
The vortex solutions that we have discussed in this section
are $1/2$ BPS objects and so preserves $6$ supercharges.
There is a subtle issue about the vortex worlsheet theory.
If the action of the $SU(2)_{C+F}$ symmetry on the supercharges
 would have been trivial, we would expect that the vortex dynamics was
 described by an $S^2$ quantum mechanics with 6 supercharges.  
Here the situation is different: only two of the six unbroken supercharges,
the ones with 
\[\omega_{12}= -\omega_{21}=\omega_{43}^*=-\omega_{34}^*=
\left(\begin{array}{c} i \\ 1\end{array}\right)\alpha_1\]
(and all the other entries $\omega_{AB}$ vanishing),
are left unchanged by a generic $SU(2)_{C+F}$ transformation. 
So  we expect that the effective quantum mechanics that describes the
vortex has only two supercharges.

A  related question is the number of
fermionic zero modes on the vortex background.
This is discussed in Appendix A for $N=2$. We find a total of eight
real fermionic zero modes is found, of which, only
six are generated by the action of the broken supercharges.
The issue of the vortex effective theory is rather tricky.
 The $\mathcal{N}=6$ mass deformed theory that we are considering
has non-central extensions in the supersymmetry algebra \cite{nahm,linma}
(which means that the anti-commutator of some of the supersymmetry generators
closes not only into a combination of translations and central charges, 
but also or R-symmetry generators).

A description of the relevant supersymmetry algebra is given in \cite{agarwal}.
Let us first introduce the mass deformed $\mathcal{N}=4$ SUSY algebra.
It consists of the Lorentz transformations $\mathcal{L}_{\alpha \beta}$,
the momentum generators $\mathcal{B}_{\alpha \beta}$, the $SU(2) \times SU(2)$ 
$R$-symmetry generators $\mathcal{R}_{ab}=\mathcal{R}_{ba}$ and
 $\dot{\mathcal{R}}_{\dot{a} \dot{b}}=\dot{\mathcal{R}}_{ \dot{b} \dot{a}}$, 
and eight supercharges $\mathcal{Q}_{\alpha b \dot{c}}$.
The anticommutator of the supercharges is:
\beq \{ \mathcal{Q}_{\alpha b \dot{c}},\mathcal{Q}_{\delta e \dot{f}} \}=
\epsilon_{be} \epsilon_{\dot{c} \dot{f}} \mathcal{B}_{\alpha \delta}-
2 m \epsilon_{\alpha \delta}  \epsilon_{\dot{c} \dot{f}} \mathcal{R}_{b e}
+ 2 m  \epsilon_{\alpha \delta}  \epsilon_{be} \dot{\mathcal{R}}_{\dot{c} \dot{f}} \, .
\eeq
The $\mathcal{N}=6$ algebra, which is the relevant one for our problem, 
 has four additional supersymmetries
$\tilde{Q}_\alpha^{\pm}$, an extra $U(1)_A$ R-symmetry $\tilde{\mathcal{B}}$
and a central charge $\tilde{\mathcal{C}}$. The non-trivial commutation relations are:
\beq [ \tilde{\mathcal{B}}, \tilde{\mathcal{Q}}_\alpha^\pm ]=
\pm \tilde{\mathcal{Q}}_\alpha^\pm  \, , 
\qquad \{ \tilde{\mathcal{Q}}_\alpha^+  , \tilde{\mathcal{Q}}_\beta^- \}=
\mathcal{B}_{\alpha \beta} -  i m \epsilon_{\alpha \beta} \tilde{\mathcal{C}} \, .\eeq
The central charge $\tilde{\mathcal{C}}$ is given by the $U(1)_B$ symmetry.

The vortex is a $\tfrac{1}{2}$ BPS objects and so we expect that it
comes in a short $\mathcal{N}=6$ multiplet \cite{beho,agarwal}, 
which consists of four bosons and four fermions.
This multiplet of eight states should be generated 
(via a Jackiw-Rebbi mechanism)
by the three complex
fermionic zero modes that correspond to broken supercharges.
The other extra complex fermionic zero mode is then interpreted
as the superpartner of the internal $S^2$ bosonic coordinate.
This shows that for $N=2$ there is no evidence of extra
bosonic internal coordinates, which (if they existed) should have had fermionic
super-partners. It might be
 that the situation changes for larger $N$, but we find this unlikely.

\section{Gravity dual of the mass deformed theory}

At any finite $k$, the ABJM theory has the interpretation of $N$
M2-branes probing a ${\mathbb C^4}/{\mathbb Z}_k$ orbifold
singularity \cite{abjm}. In the large $N$, strong coupling limit, this
allows to identify the gravity dual as eleven dimensional
supergravity on $AdS_4 \times S^7/{\mathbb Z}_k$, with $N$ units of
four-form flux. Viewing the $S^7$ as a Hopf fibration of $S^1$ over
${\bf CP}^3$, at large $k$, a reduction to type IIA string theory
becomes possible. Then the gravity dual of the ABJM theory in the 't
Hooft large $N$ limit, as $k\to \infty$, 
with $\lambda=N/k$ fixed and large, is the type IIA string theory on
$AdS_4 \times {\bf CP}^3$ \cite{abjm}. The background has $N$ units of
Ramond-Ramond four-form  flux on $AdS_4$ and $k$ units of two-form
flux on a ${\bf CP}^1\subset {\bf CP}^3$. 

We will adopt a similar approach to obtain the gravity dual of the
mass deformed ABJM theory. This is a two step process. First we recall
the results of Lin, Lunin and Maldacena (LLM) \cite{llm} and those of
Bena and Warner \cite{bena}, for the mass
deformation of the theory on $N$ M2-branes probing flat space. The 
large $N$ gravity dual of that theory (with a large set of vacuum
states) is given by the $SO(4) \times SO(4)$ symmetric LLM solutions of
\cite{llm}. We will take this gravity solution and perform a
${\mathbb Z}_k$ quotient on it to yield the mass deformed ABJM theory
for generic $k$. Subsequently we will reduce this to a type IIA
solution in the limit of large $k$ and investigate the dynamics of
vortices in this strongly coupled description.

\subsection{The background for $k=1$ }
For $k=1$, which is the mass deformed theory on a large $N$ number of
M2-branes, the dual eleven dimensional metric in the notation 
of \cite{llm}, takes the form
\begin{eqnarray}
&& ds^2_{11}= e^{4\tilde \Phi/3} (-dt^2+dw_1^2+dw_2^2) +
e^{-2\tilde\Phi/3}\left(\,h^2\,(dx^2+dy^2) + y e^G\,d\Omega_3^2+ y
  e^{-G} \,
d\tilde{\Omega}_3^2 \right)\, ,
\nonumber\\
&&e^{2\tilde \Phi} = {1\over h^2- V_1^2(x,y)/h^2}\,,\qquad
\frac{1}{h^2}= 2y \cosh G\,, \qquad 2 \,z(x,y) = \tanh G.
\label{massdef}
\end{eqnarray}
The functions $z$ and $V_1$ on the $x-y$ plane are specified by a
choice of the positions of M5-branes wrapping one or the other of the
two $S^3$'s in the geometry. The distribution of wrapped M5-branes
picks out a particular vacuum of the mass-deformed M2-brane
theory. The wrapped M5's arise as usual due to the deformation
which blows up multiple M2-branes into fivebranes.

Let us make a few technical remarks in order to make contact 
with the notation used in \cite{bena} by Bena and Warner.
In \cite{bena}, the coordinates $(x,y)$ are replaced by $(u,v)$.
The relation between the two choices of variables is the following:
\beq x=4 L^2 (u^2-v^2) \, , \, \qquad y=8 L^2 u v \, ,\eeq
where $L$ is a constant that
 corresponds to the scale of the mass deformation.
Further, the solution in \cite{bena} is given in term of a
function $g(u,v)$; the relation between $g$ and the
function $z(x,y)$ used above \cite{llm} is,
\beq \partial_x g = -\frac{1}{4} \left(z-\frac{x}{2 \sqrt{x^2+y^2}}
\right) \,.
\label{gdef}
\eeq 
Finally, the constant $\beta^2$ in \cite{bena}
has to be set equal to $1/8$ in order to obtain the non-singular solutions
discussed in \cite{llm}.

For the sake of completeness let us also write down the three form potential
in this background,
\beq
C_3= - \frac{e^{2\tilde \Phi}V_1}{h^2} dt\wedge dw_1\wedge dw_2
+\mathcal{A} \, d\Omega_2 \wedge
  \left(d\lambda+d\varphi\right) + \mathcal{B} \, d\tilde\Omega_2\wedge 
\left(d\lambda-d\varphi\right) \, .
\eeq
The functions $\mathcal{A}$ and $\mathcal{B}$ 
are then
more straightforward to write in the Bena-Warner notation \cite{bena},
\beq \mathcal{A}= \frac{1}{\beta}
\left( g-\frac{L^2 u (u^2+v^2) (\partial_u g) }
{2 L^2 v^2 - v (\partial_v g)+ u (\partial_u g) }\right) \, ,
\eeq
\[
\mathcal{B} = - \frac{1}{\beta} 
\left( g-\frac{L^2 v (u^2+v^2) (\partial_v g) }
{2 L^2 u^2 + v (\partial_v g)- u (\partial_u g) }\right) \, ,
\]
where $\beta=1/\sqrt{8}$.

%\subsubsection{The Higgs vacuum}
The vacuum of the mass-deformed M2-brane theory is specified by
the choice of the functions $z(x,y)$ and $V_1(x,y)$. In particular,
since the Higgs vacuum in the field theory corresponds to an
irreducible representation for the $N\times N$ matrices giving VEVs to
the bifundamental matter fields, we expect that there is a single
dielectric M5-brane made from blowing up the $N$ M2-branes. In the
large $k$ limit, where the semiclassical analysis of the ABJM theory
holds, we saw a fuzzy two sphere structure \cite{nastase} which can be interpreted as
$N$ D2-branes polarized into a single wrapped D4-brane in type IIA theory. When
lifted to M theory this becomes a single M5-brane. In the free fermion
picture of \cite{llm}, this is represented as in Fig.~2 
as a black strip, corresponding to a highly energetic particle
state. The position of the strip on the $x$-axis and its width are
dictated by the number of M2-branes and the number of wrapped M5's.
\begin{figure}[h]
\begin{center}
\epsfig{file=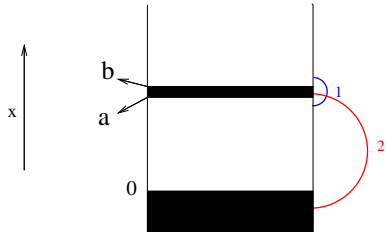,width=2.0in}
\end{center}
%\noindent\small{\bf Figure 2:} 
\caption{The Higgs vacuum is given by one
  wrapped dielectric M5-brane which translates to a highly energetic
  particle in the fermion fluid picture.}
\label{onestrip}
\end{figure}

The strip in Figure~2 represents a section of the geometry
at $y=0$. The vertical axis is the coordinate $x$. In the black region
the first sphere $S^3$ shrinks to zero size. Similarly, in the white region
the second sphere $\tilde{S}^3$ shrinks to zero. 
At the boundary of the white and the black
regions, both the three-spheres shrink.
Let us denote with $(a,b)$ the position of the lower and upper bounds
of the black strip  in Figure 2.
In term of the function $z(x,y)$ this means that,
\begin{eqnarray}
z(x,y=0) = &&\frac{1}{2}\,,\quad{\rm for}\quad{0 < x < a}
\quad{\rm and}\quad{x > a+b} \, , \\\nonumber
-&&\frac{1}{2}\,,\quad{\rm for}\quad {a < x < a+b } \quad{\rm
  and}\quad{x < 0} \, .  
\end{eqnarray}
We can consider arcs in the $(x,y)$ plane that enclose a black or 
a white strip (see for example the arcs 1 and 2 in Figure~2) and 
construct a four-sphere by taking one of these arcs and tensoring with the $S^3$
that shrinks to zero at the tips of the arc. The flux of $F_4$
over each of these four-spheres is equal to the thickness of the
strip enclosed by these arcs. For this reason the thickness of each strip
must be an integer.

 The fluxes on the four-spheres that are enclosed by
the arcs 1 and 2 are proportional to $(b-a)$ and $a$, respectively.
If $(b-a)<<a$, we may think of the first $S^4$ (constructed using arc
1) as being 
transverse to the M5-branes. Then $(b-a)$ corresponds 
to the number of M5-branes which
are blowing up on a three-sphere.
The second $S^4$ arises in the following way. Let us consider
the three-sphere that the M5-branes are wrapping. At the center of the space
this three-sphere is contractible. As we move away from the center
towards the M5's, 
the backreaction of the branes on the geometry makes the $S^3$ contract
again.  This produces the $S^4$ which is enclosed by the arc 2 in Figure 2.
The product of the two $F^4$ fluxes is the total 
M2-brane charge.

The Higgs vacuum configuration is given by the following
solutions for $z$ and $V_1$,
\begin{eqnarray}
&&z = \frac{1}{2}\left(\frac{x}{\sqrt{x^2+y^2}}-
\frac{x-a}{\sqrt{(x-a)^2+y^2}}+
 \frac{x-b}{\sqrt{(x-b)^2+y^2}}\right)\, ,\\\nonumber
&&V_1=\frac{1}{2}\left( \frac{1}{\sqrt{x^2+y^2}}-
\frac{1}{\sqrt{(x-a)^2+y^2}}+\frac{1}{ \sqrt{(x-b)^2+y^2}}\right)\,,
\quad a= N'\,,b= N'+1,
\end{eqnarray}
where $N'$ is the M2-brane charge.
In the notation of \cite{bena} where the solutions are written in
terms of the function $g$,  (see Eq.\eqref{gdef}),
\beq g=\frac{-\sqrt{(x-b)^2+y^2} +\sqrt{(x-a)^2+y^2}}{8} \, .\eeq

We can also consider more general solutions,
with an arbitrary number of black strips,
\begin{eqnarray}
&&z = \frac{1}{2}\left(\frac{x}{\sqrt{x^2+y^2}}+ \sum_i
 \frac{x-b_i}{\sqrt{(x-b_i)^2+y^2}}
-\frac{x-a_i}{\sqrt{(x-a_i)^2+y^2}}\right)\, ,\\\nonumber
&&V_1=\frac{1}{2}\left( \frac{1}{\sqrt{x^2+y^2}}
+ \sum_i \frac{1}{ \sqrt{(x-b_i)^2+y^2}}
- \frac{1}{\sqrt{(x-a_i)^2+y^2}}+\right)\, ,
\end{eqnarray}
where $(a_i,b_i)$ are the positions of the lower and upper bounds
of each of the strips.
On the field theory side, these correspond to other vacua of the theory.
The full set of strip configurations 
with a fixed M2-brane charge $N'$, 
can be classified by Young Tableau with $N'$ boxes.
Their total number is given by the
number partitions of $N'$.
As pointed out in \cite{massivo}, there is a mismatch
between this and the number of vacua in the {\em classical} field theory.
The solution to this puzzle is still unknown.
It is possible that this is due to the fact that not all vacua
of the theory can be realized within the supergravity approximation. 
Another option is that quantum effects may possibly break supersymmetry
in some of the classically visible vacua of the mass-deformed ABJM
theory.

%The one other feature of the metric \eqref{elevend} is that along the
%$x$-axis, one of the two $S^2$'s are always of zero size. In
%particular we have that
%\begin{eqnarray}
%z(x,y=0) = &&\frac{1}{2}\,,\quad{\rm for}\quad{0\leq x < Nk}
%\quad{\rm and}\quad{x > N k+1}\\\nonumber
%-&&\frac{1}{2}\,,\quad{\rm for}\quad {Nk < x < N k+1}.
%\end{eqnarray}
%It follows then that within the interval $Nk < x< Nk+1$, the first
%$S^2$ is of zero size while $\tilde S^2$ shrinks along the rest of the
%positive $x$-axis. 

\subsection{${\mathbb Z}_k$ quotient and reduction to type IIA}

In this section we perform a $\mathbb{Z}_k$ quotient of the
$k=1$ solution. In order to keep the number of M2-branes fixed and equal to $N$,
we have to set $N'=k N$.  

Let us parameterize the eight directions transverse to the M2-branes, in
terms of the  four complex coordinates $z_i$, ($i=1,\ldots 4$),
\begin{eqnarray}
&&
z_1= u \, \sin \eta \, 
e^{{i}(\lambda+\theta+\varphi)} \, ,
\qquad
 z_2=u \,
\cos \eta \,e^{i(\lambda-\theta+\varphi)}\, ,\\\nonumber
&& z_3= v \,\sin \tilde{\eta} \, e^{i(-\lambda+\tilde{\theta}+\varphi)} \, ,
\qquad z_4= v \,
\cos \tilde{\eta} \, e^{i(-\lambda-\tilde{\theta}+\varphi)}\, .
\end{eqnarray}
% and the $x,y$ coordinates are given by,
% \beq x= r \cos 2\psi \, \qquad y= r \sin 2\psi \, .\eeq
In this parametrization, the metrics for the two three-spheres in the 
eleven dimensional background
\eqref{massdef} are,
\begin{eqnarray}
&&d\Omega_3^2 = d\eta^2 +\sin^2 2\eta\,d\theta^2+
((d\lambda+d\varphi)-\cos 2\eta\,d\theta)^2,
\\\nonumber\\\nonumber
&&d\tilde\Omega_3^2 = d\tilde\eta^2 +
\sin^2 2\tilde\eta\, d\tilde\theta^2+
((d\lambda-d\varphi)+\cos 2\tilde\eta\,d\tilde\theta)^2.
\end{eqnarray}
So each $S^3$ is viewed as a Hopf fibration of an $S^1$ 
over $S^2$, and the background has an $SO(4)\times SO(4)$ isometry,
acting naturally on the three-spheres. The mass deformed ABJM theory
should only retain an $SU(2)\times SU(2)\times U(1)\times U(1)$
isometry. 
This can be achieved by an appropriate quotient action
on a linear combination of the two $S^1$'s, namely the $\varphi$
coordinate.  
The ${\mathbb Z}_k$ quotient we perform, acts on the coordinates as
\beq
z_j \rightarrow z_j \,e^{i\frac{2\pi}{k}}\,,
\qquad \varphi\rightarrow\varphi+2\pi/k.
\eeq
Hence in the limit $k\rightarrow\infty$, the period of the 
 angular coordinate $\varphi$
shrinks and we may pass to the weakly coupled type IIA description. To
implement this, it is useful to first perform a rescaling 
$\varphi\rightarrow \varphi/k$, and then write the eleven dimensional
metric as 
\begin{eqnarray}
ds^2_{11} = &&e^{4\tilde \Phi/3}(-dt^2+ dw_1^2+dw_2^2)+e^{-2\tilde\Phi/3}
\left[h^2(dx^2+dy^2) + y e^G(d\eta^2+\sin^22\eta d\theta^2)\right.
\nonumber\\\label{elevend}\\\nonumber
&&\left.
+y e^{-G}(d\tilde\eta^2+\sin^22\tilde\eta\, d\tilde\theta^2)
 + \frac{2y}{\cosh G}\left(d\lambda-\frac{1}{2}\cos 2\eta\, d\theta
+\frac{1}{2}\cos 2\tilde \eta \,d\tilde \theta
\right)^2 \right.
\\\nonumber
&&\left. + 2 y \cosh G \, \frac{1}{k^2}(d\varphi + k\,\omega)^2\right],
\end{eqnarray}
Here $\varphi$ has period $2\pi$ and $\omega$ is the one-form,
\beq
\omega = \tanh G\,d\lambda - \frac{e^G}{2\cosh G} \cos2\eta\,d\theta
- \frac{e^{-G}}{2\cosh G}\cos 2\tilde \eta\, d\tilde\theta.
\label{1form}
\eeq
This metric has the manifest $SU(2)\times SU(2)$ isometry of the two
spheres, and the two $U(1)$ isometries corresponding to shifts of
$\varphi$ and $\lambda$. 
We see below that when we focus on a specific vacuum of the mass
deformed theory, the resulting metric asymptotes to $AdS_5 \times
S^7/{\mathbb Z}_k$ as it should.

With this choice of the vacuum we can now 
determine some features of the geometry including the
large $r=\sqrt{x^2+y^2}$ asymptotics. As $r \rightarrow \infty$, we find
\beq
e^{-2\tilde\Phi}\simeq \frac{Nk}{r^3}\,,\qquad
h^2\simeq \frac{1}{2r}\,,\qquad e^G \simeq \cot\psi,
\eeq
so that the metric asymptotes to $AdS_4\times S^7/{\mathbb Z}_k$
\beq
ds^2_{11}\simeq\frac{r^2}{(N k)^{2/3}}(-dt^2+dw_1^2+dw_2^2)+
(Nk)^{1/3}\frac{dr^2}{2 \,r^2}+ 2 (Nk)^{1/3}ds^2_{S^7/{\mathbb Z}_k}.
\eeq
Subsequent reduction to type IIA in the large $k$ limit will give the 
$AdS_4\times{\bf CP}^3$ background of \cite{abjm}.

Let us quickly sketch how to pass from the
eleven dimensional description to the 10-dimensional 
type IIA one. Writing the metric as
\beq ds^2=G_{mn}^{10} dx^m dx^n + e^{2 \gamma} (dx^{11} -A_m dx^m)^2 \, ,\eeq
then the scalar $e^{3 \gamma}$ is proportional  to the string theory
dilaton $e^{2 \phi}$.  Comparing with our eleven dimensional
background, we conclude that
\beq
e^\phi = e^{-\tilde\Phi/2}\,\left(k \,h \right)^{-3/2}.
\eeq
It is easy to check that the dilaton is bounded and therefore small
everywhere,  for large enough $k$. In addition, the dilaton vanishes
at $x=a=Nk$ and $x=b=Nk+1$.

Finally, we can write the string frame metric as, 
\begin{eqnarray}
ds^2_{\rm string}= &&e^{2\phi/3}\,G^{10}_{mn}dx^m
dx^n\label{2ametric}\\\nonumber 
&& =e^{\tilde\Phi}(hk)^{-1}\left(-dt^2+dw_1^2+dw_2^2\right)+
e^{-\tilde\Phi}(hk)^{-1}\left[h^2(dx^2+dy^2) +
\right.\\\nonumber\\\nonumber 
&&\left.y e^G(d\eta^2+\sin^22\eta d\theta^2)
+y e^{-G}(d\tilde\eta^2+\sin^22\tilde\eta\, d\tilde\theta^2)
 + \right. \\\nonumber\\\nonumber
&&\left.\frac{2y}{\cosh G}\left(d\lambda-\frac{1}{2}\cos 2\eta\, d\theta
+\frac{1}{2}\cos 2\tilde \eta \,d\tilde \theta
\right)^2 \right].
\end{eqnarray}
and the Ramond-Ramond one-form potential $C_1$
\beq
C_1= k\,\omega
\eeq
where $\omega$ is the one-form defined in Eq.\eqref{1form}. The 
type IIA background will also have a $B_2$ Neveu-Schwarz potential 
switched on and a three-form Ramond-Ramond potential originating from
the eleven dimensional three form $C_3$. We will not need these for
our analysis of the dynamics of the probe D0-brane which is identified
as the vortex soliton of the mass deformed ABJM theory.

\subsection{Probe D0-brane dynamics}

The vortex soliton in the mass deformed ABJM theory carries a charge
which is an integer multiple of $k$, under the $U(1)$ symmetry
generated by $^*F_{\tilde b}$. On the string theory side, this
symmetry is generated by,
\beq J = k \, Q_0 + N \, Q_4 \, ,\eeq
where $Q_0$ and $Q_4$ are the D0-brane and the D4-brane charges.
Hence it is natural to identify the vortices
(which indeed carry $k$ units of $J$ charge, as we saw in the field
theory) with the D0-branes. In the type IIA brane picture, we expect
that the mass-deformed ABJM theory (for $k\gg 1$) is realized on 
dielectric D4-branes arising from a blown-up configuration of
D2-branes. A D0-brane can form a bound state with the dielectric 
D4-brane and appear as a vortex soliton in the three dimensional gauge 
theory\footnote{This picture is rather similar to that of flux tubes and
vortex strings in ${\cal N}=1^*$ theory \cite{Polchinski, us}, which
arise from F1/NS5 and D1/D5 bound states.}.

The action for a probe D0-brane is given by the sum of the Born-Infeld
and of the Chern-Simons term,
\beq 
S_{\rm D0}=\int d \xi^a (e^{-\phi} \sqrt{-G_{aa}} +C_a) \, . \eeq
Let us first consider, a time independent probe D0-brane.
Then the only contribution 
to the action comes from the Born-Infeld term.
We identify this with the mass of the vortex 
\beq m = k\, e^{\tilde\Phi}\,h
%k
%\frac{(1-4 z^2)^{1/2}}{2  v  (h^4-V_1^2)^{1/2} }
= k \frac{1}{\sqrt{1-V_1^2/h^4}}\, . \label{minimize} \eeq
This quantity is minimized when
\beq V_1=\frac{1}{2}\left( \frac{1}{\sqrt{x^2+y^2}}-
\frac{1}{\sqrt{(x-a)^2+y^2}}+\frac{1}{ \sqrt{(x-b)^2+y^2}}\right) = 0 \, , \label{mspace}\eeq
and the value of the soliton mass is
\beq m = k \, .\eeq
This matches with the value computed for the 
mass of the vortex soliton in Section 3
(in our string theory calculation we are working in the
dimensionless units with $\mu=1$).

\subsubsection{Probe moduli space}
It is fairly clear from \eqref{mspace}, 
that the probe action attains its minimum value
along a one dimensional curve in the $(x,y)$ plane.
The moduli space for the probe D0-brane is therefore  a
{\em six dimensional} manifold $\mathcal{P}$ obtained by 
$S^2\times \tilde S^2\times S^1$
fibred along the one dimensional curve given by Eq.\eqref{mspace}, 
where the $S^1$ is also non-trivially fibred over the two $S^2$'s.
The shape of $\mathcal{P}$
projected onto the $(x,y)$ plane is shown in Figure 3.
\begin{figure}[h]
\begin{center}
\epsfig{file=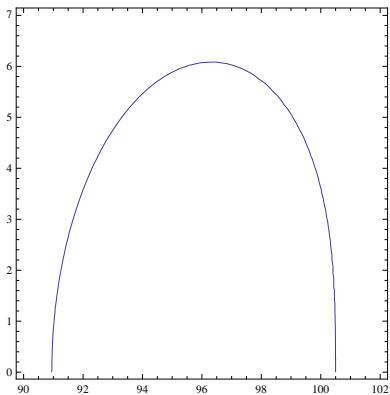,width=2.0in}
\end{center}
%\noindent\small{\bf Figure 3:} 
\caption{The curve  $V_1=0$ in the $(x,y)$ plane, where the D0-brane action is
minimized. In this plot we have used the numerical values $a=100$, $b=101$.
At the points where the curve intersects the $x$-axis, one of the two
$S^2$'s and an 
$S^1$ shrink to zero size. The point near $x=91$ corresponds to the
vortex solution visible in the field theory. }
\label{vv1} 
\end{figure}
For each value of $x$, with
\beq  
\tilde{x}_1 \leq x \leq \tilde x_2\,;\qquad 
\tilde x_1= b- \sqrt{b^2- a b}\,,\quad\tilde x_2= \sqrt{a b}\, , \eeq
 there exists only one solution to Eq.(\ref{mspace}).
Denoting this solution as $\tilde{y}(x)$, we may
consider sections of $\mathcal{P}$ at constant $x$.
for  generic $\tilde{x}_1<x<\tilde{x}_2$ the section
is five dimensional and can be parameterized with the 
five coordinates $(\eta,\theta,\tilde{\eta},\tilde{\theta},\lambda)$.

The topology of a cross section at a generic point of the segment with $\tilde{x}_1<x<\tilde{x}_2$
is equivalent to $S^3 \times S^2$.
When $x=\tilde{x}_1$  the  $S^3$ obtained by fibering
$S^1$ over  $\tilde{S}^2$ 
shrinks to zero size. This 
section of the moduli space is parameterized  by the $S^2$ coordinates
$(\eta,\theta)$. 
At $x=\tilde{x}_2$, 
the  $S^3$ obtained by fibering
$S^1$ over  $S^2$  shrinks to zero
 and the section is parametrized 
by the $\tilde{S}^2$ coordinates $(\tilde{\eta},\tilde{\theta})$.

The solitonic vortex solution that we have found in the weakly coupled limit
in Section 3 maps to the probe D0-brane at $x=\tilde{x}_1$ at strong
coupling.  At
this special point the $S^2$ is finite sized. The dielectric D4-brane 
wraps this $S^2$ and the probe D0-brane spontaneously breaks the
associated $SU(2)$ isometry. The position of the D0-brane 
on this $S^2$ corresponds to the internal 
orientation of the vortex in the colour-flavour space. At this point,
the shrunk $\tilde S^2$ and $S^1$ imply that the vortex solution
explicitly preserves an $SU(2)\times U(1)$ global symmetry. The
unbroken $SU(2)$ can be identified as the symmetry that acts on the
doublet $(R^1,R^2)$ in the field theory.

%The moduli space $\mathcal{P}$ that we have found at strong coupling
%within the supergravity approach is then bigger than the space
%of classical vortex solutions that we have found at weak coupling. 
%A possible explanation for this issue could be that 
%we have not found all the classical vortex solutions at weak
%coupling. It is possible that our weak-coupling  
%ansatz was not enough to accomodate the most general vortex solution.
%In particular, in our classical solitonic solution we have always kept the
%scalars $R^\alpha =0$. The presence of extra directions in the moduli
%space at strong coupling, where $\tilde S^2$ is blown up, suggests
%that these correspond to solutions where the $R^\alpha$ are turned on.
%It is possible that more general solutions with
%$R^\alpha \neq 0$ exist at weak coupling, 
%although these may not be realized in a
%simple way since the Higgs vacuum should have 
%$R^\alpha=0$ asymptotically. 
%We leave this question as a topic for future investigation.  

\subsubsection{Moduli space effective action}

From the probe D0-brane action it is straighforward to
find the vortex effective theory. The bosonic part of the vortex
quantum mechanics is a 
1-dimensional sigma model with target space $\mathcal{P}$,
which can be parameterized by the five coordinates
$(x,\eta,\theta,\tilde{\eta},\tilde{\theta},\lambda)$
(the value of $y=\tilde{y}(x)$ can be found by inverting
Eq.~(\ref{mspace})). Allowing a slow time dependence for the vortex
position in ${\cal P}$, the the D0-brane action
can be expanded out up to second order in time derivatives
\beq
S_{D0}\big|_{\cal P}= k + S_1 +S_2,
\eeq
where the first contribution is the D0-brane/vortex mass, and $S_1,
S_2$ are the first and second order derivative terms respectively. 
The moduli space metric can be
read from the Born-Infeld part of the action, while 
the first order terms follow from the coupling of  the D0 to the
Ramond-Ramond one-form, $C_1 = k\omega$.
The second order kinetic terms are 
\begin{eqnarray}
S_2 = && 
\frac{1}{2}
\int dt \left[
a_1 \left(\dot{\eta}^2+(\sin^2 2 \eta) \, \dot{\theta}^2\right)+
a_2 \left(\dot{\tilde{\eta}}^2+(\sin^2 2 \tilde{\eta}) \, \dot{\tilde{\theta}}^2 \right)+ \right.
\\\nonumber
&&  \left. + a_3 \left( 1+ \left(\frac{d \tilde{y}}{d x}\right)^2\right)
\dot{x}^2 + a_4 \left( \dot{\lambda} -\frac{\cos 2 \eta}{2} \dot{\theta} + 
\frac{\cos 2 \tilde{\eta}}{2}  \dot{\tilde{\theta}}\right)^2 
\right] \,  \, .
\end{eqnarray}
The coefficients of the second derivative terms evaluated on the
moduli space, are
\begin{eqnarray}
&&a_1=\frac{1}{2}k \,(1+ 2z)\big|_{\cal P}\, ,\qquad a_2= 
\frac{1}{2}k\,(1-2 z)\big|_{\cal P}
\\\nonumber
&& a_3 =\frac{k}{4y^2} \,(1-4 z^2)\big|_{\cal P} \, , \qquad
a_4= k \, (1-4z^2)\big|_{\cal P}.  
\end{eqnarray}
A  numerical plot of the functions $a_j$ is given in Figure 4 and
5. The six dimensional moduli space is a deformation of ${\bf CP}^3$,
preserving an $SU(2)\times SU(2)\times U(1)$ isometry.

\begin{figure}[h]
\begin{center}
\epsfig{file=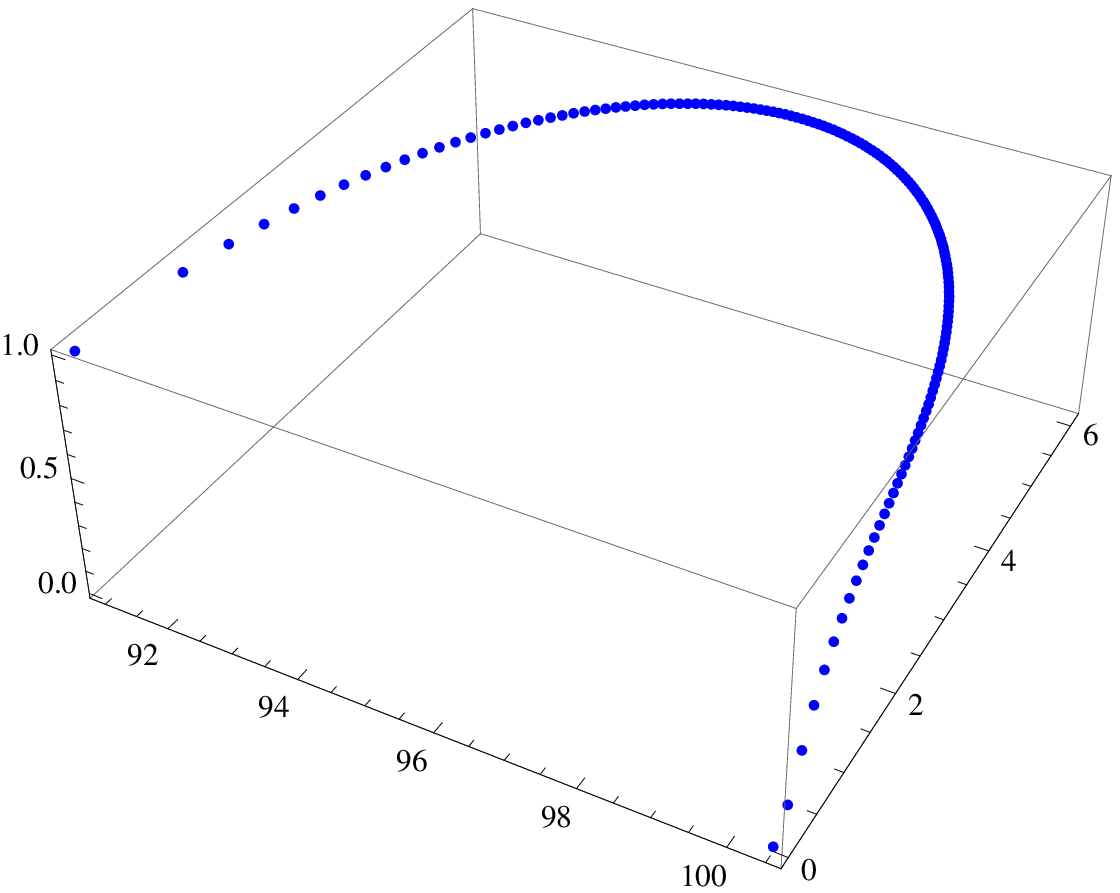, width =2.2in}\hspace{0.5 in}
\epsfig{file=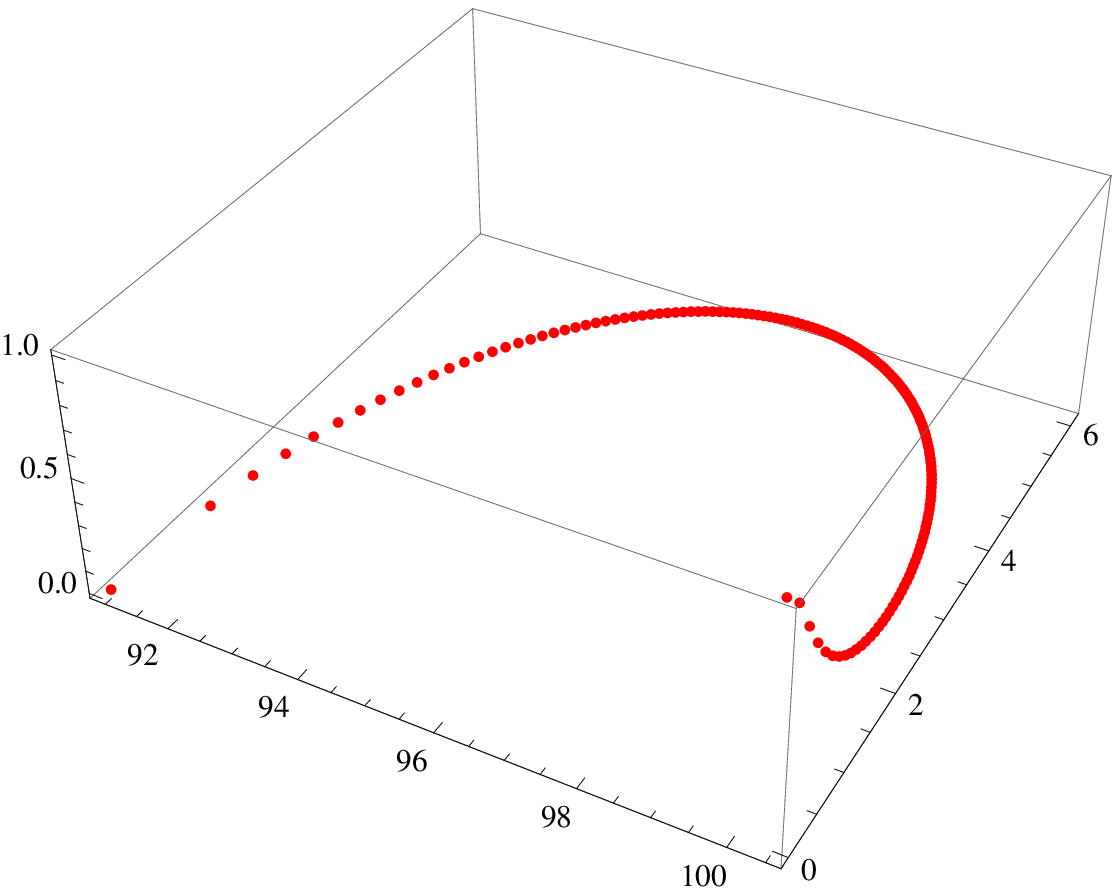, width=2.2in}
\end{center}
%\epsfxsize=6cm \centerline{\epsfbox{nc2.eps}}
%\noindent\small{\bf Figure 4:} 
\caption{Kinetic terms $a_1$, $a_2$ (in units of $k$)
for each of the sphere components $S^2$, $\tilde{S}^2$
 as a function of $(x,y)$.}
\end{figure}
\begin{figure}[h]
\begin{center}
\epsfig{file=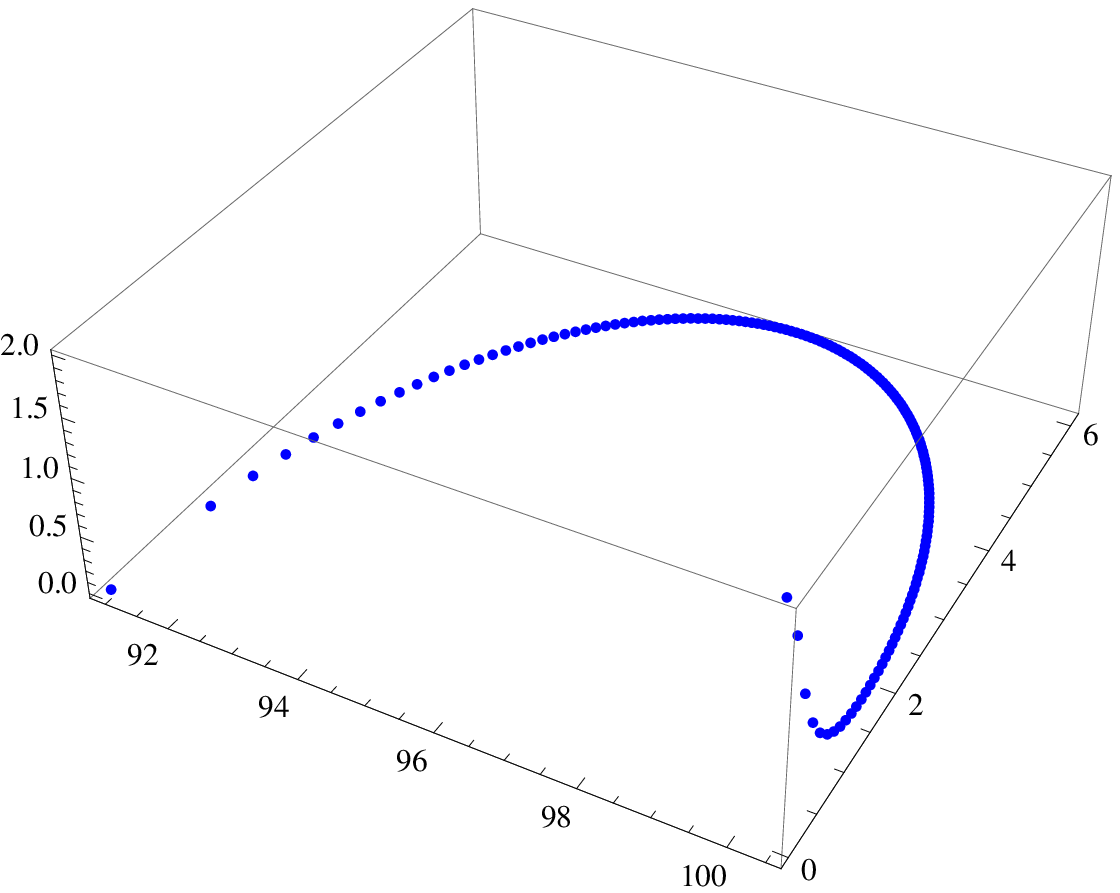, width =2.2in}\hspace{0.5 in}
\epsfig{file=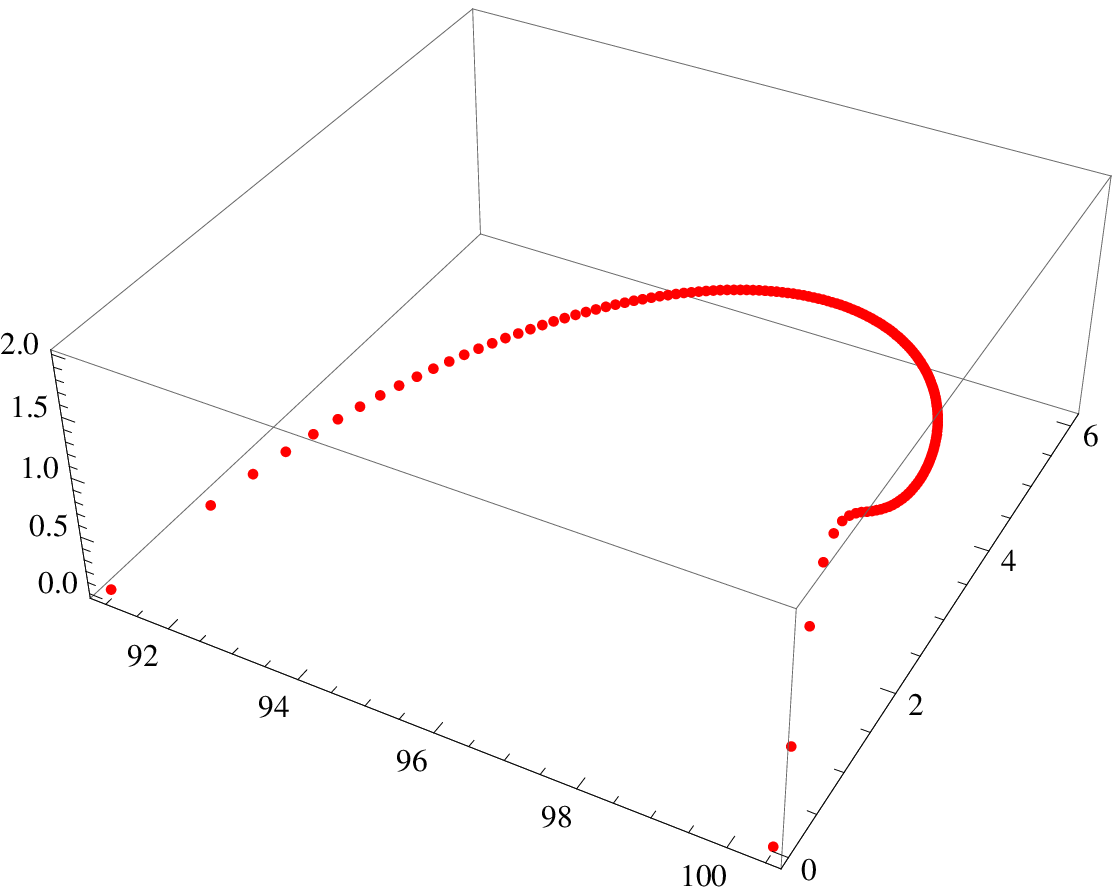, width=2.2in}
\end{center}
%\epsfxsize=6cm \centerline{\epsfbox{nc2.eps}}
%\noindent\small{\bf Figure 5:} 
\caption{Kinetic terms $a_3$, $a_4$ (in units of $k$)
 as a function of $(x,y)$.}
\end{figure}

The first order terms in the D0-brane action are
\beq 
S_1=\int dt \left( a_1  \cos 2\eta\,\dot{\theta} + a_2 \cos 2\tilde
  \eta\, \dot{\tilde\theta} + (a_1-a_2) \dot{\lambda}  
 \right) \, ,\eeq
%where
%\beq b_1=k \, \frac{1+ 2 \, z}{2} \, \qquad b_2=k \, \frac{1- 2 \,
%z}{2} \,  \, \qquad  
%b_3=k \, 2 z  \, .\eeq
%A numerical plot of $b_j$ is given in Figure 6.
which describe the motion of the particle in the presence of $k$ units
of magnetic flux through $S^2$ and $\tilde S^2$.

We may add a total derivative term to the action and put it in a form
where the physical interpretation becomes manifest,
\beq
S_1=\int dt \left( (a_1 \, \cos 2\eta-1)\,\dot{\theta} + 
(a_2 \cos 2\tilde\eta-1)
\, \dot{\tilde\theta} + (a_1-a_2) \dot{\lambda}  
 \right) \,.
\eeq 
Now, it is interesting to look at this action
at the point in the $(x,y)$ plane  that naturally corresponds to the $S^2$ 
moduli space of vortex solitons that we have found at weak coupling.
This is the point $(x,y)=(\tilde{x}_1,0)$ or equivalently $z=\frac{1}{2}$.
At this point where $\tilde S^2$ vanishes, the action is precisely that
of a particle moving on $S^2$ with radius $\sqrt{k/2}$, 
in the presence of a Dirac monopole
connection of strength $k$,
\beq
L_{\rm vortex}\big|_{z=\frac{1}{2}}= 
\frac{k}{2}\left[\frac{1}{2}(\dot\eta^2 + \sin^2 2\eta\,\dot\theta^2)+
  (\cos2\eta-1)\dot\theta\right].
\eeq
Note that this is the Dirac monopole connection on the ``north pole''
patch, and is singular at the south pole.
This is similar to the non-Abelian Chern-Simons vortex discussed in \cite{ct}.
At the classical level, it appears consistent to identify this as the
moduli space action for the vortex soliton we found at weak coupling
since it preserves the same symmetries.
%\begin{figure}[h]
%\begin{center}
%\epsfig{file=bplot1.eps, width =2.2in}\hspace{0.5 in}
%\epsfig{file=bplot2.eps, width=2.2in} \hspace{0.5 in}
%\epsfig{file=bplot3.eps, width=2.2in} \hspace{0.5 in}
%\end{center}
%\epsfxsize=6cm \centerline{\epsfbox{nc2.eps}}
%\noindent\small{\bf Figure 6:} 
%\caption{First order terms $b_1$, $b_2$, $b_3$ (in units of $k$)
% as a function of $(x,y)$.}
%\end{figure}
%In correspondence of this point $a_1=k$ and $a_{2,3,4}=0$;
%the moduli space metric is then a sphere with radius
%proportional to the Chern-Simons level $k$.
It is interesting that the radius of this sphere is quantized and
 determined by the Chern-Simons level $k$. This also appears to be
 manifest at weak coupling where the radius of the fuzzy two-sphere
 in the Higgs vacuum, in Eq.\eqref{fuzzyeq}, after dividing out by a
 factor of $N$ to normalize the corrdinates, is proportional to
 $\sqrt k$.
%Also $b_1=k$ and $b_2=0$; this correponds to 
% the magnetic monopole
% gauge connection  (with magnetic charge $k$)
%on the $S^2$ moduli space. 

\section{Discussion and Conclusions}

In this paper, we found $\frac{1}{2}$-BPS vortex solitons in the
${\cal N}=6$ mass deformation of ABJM theory. We verified that they
preserve six supercharges. These vortices in the
Higgs vacuum have internal, non-Abelian, orientational collective
coordinates which are responsible for a ${\bf CP}^1$  moduli space of
solutions. We also obtained the strong coupling gravity dual of the
mass deformed theory and its Higgs vacuum, by performing a $\mathbb{Z}_k$ quotient on the
solution found in \cite{bena,llm}. Probe D0-branes in this
background correspond to the Chern-Simons vortices. We found that the
probe D0-brane exhibits a much larger moduli space than expected for
the classical vortex soliton. Within this larger moduli space we could
however identify a section which coincides with the classical moduli space of
solutions originating from the breaking of a colour-flavour locked
symmetry.  The dynamics on this section is that of a point particle
moving on a sphere of radius $\sqrt{k/2}$ coupled to a Dirac monopole
field of strength $k$. 

The enlarged moduli space ${\cal P}$ at strong coupling leaves us with a
puzzle. We think that the extra four dimensions in the moduli space are 
an artifact of the strong coupling limit and of the supergravity approximation.
Another possible explanation could be that we have not found the
most general vortex solution because our ansatz was not sufficiently general.
We believe that the latter explanation is unlikely - it appears
difficult to arrive at a reasonable ansatz that could realize this
possibility. Also, as discussed in Section (\ref{caporetto}), the number of fermionic 
zero modes (which has been computed for $N=2$ in Appendix A) does not suggest
the existence of extra bosonic zero modes. Another 
particularly interesting feature of our solution, which introduces
further subtleties,  is that the
colour-flavour locked symmetry actually involves a locking between an
$SU(2)$ R-symmetry and the global gauge rotations. This is unusual in
that although a static vortex solution preserves six supercharges, an
adiabatic variation of the internal orientational modulus preserves
only two supercharges. The full implications of this for the vortex
effective theory also need to be understood.

{\bf Acknowledgements:} We would like to thank Tim Hollowood and Dave Tong 
for discussions.

\startappendix
\Appendix{Fermionic zero modes for $N=2$}

In this appendix we compute the number of fermionic zero modes on the vortex background
 for $N=2$. The equations for the 
two sectors $\xi$ and $\chi$  decouple from each other
and can be analyzed separately. 
We find four real zero modes in each sector.

\subsection{$\xi$ sector}
On the vortex background, the fermionic part of the action 
for this sector can be written as:
\beq -i \, \Tr (\xi^\dagger)^I \gamma^\mu D_\mu \xi_I     
 +\frac{2 \pi i}{k} \, \Tr \left(
(-1 \frac{k \mu}{2 \pi} -Q_1^\dagger Q_1 + Q_2^\dagger Q_2 ) (\xi_1^\dagger \xi_1)+ \right. \eeq
\[ \left. +
(-1 \frac{k \mu}{2 \pi} +Q_1^\dagger Q_1 - Q_2^\dagger Q_2 ) (\xi_2^\dagger \xi_2)
 -2 Q_1^\dagger Q_2 (\xi_1^\dagger \xi_2) -  2 Q_2^\dagger Q_1 (\xi_2^\dagger \xi_1) + \right. \]
\[ \left. + \xi_1^\dagger (Q_1 Q_1^\dagger-Q_2 Q_2^\dagger ) \xi_1 -
 \xi_2^\dagger (Q_1 Q_1^\dagger-Q_2 Q_2^\dagger ) \xi_2 + 
  2 \xi_1^\dagger (Q_2 Q_1^\dagger ) \xi_2 + 2 \xi_2^\dagger (Q_1 Q_2^\dagger ) \xi_1
\right) \, .\]
The following Dirac equations are found:
{ \small
\[ -\gamma^\mu D_\mu \xi_1 + \frac{2 \pi}{k} \left(
\xi_1 (-1 \frac{k \mu}{2 \pi} -Q_1^\dagger Q_1+Q_2^\dagger Q_2) - 2  \xi_2 Q_1^\dagger Q_2 
+(Q_1 Q_1^\dagger- Q_2 Q_2^\dagger) \xi_1 + 2 Q_2 Q_1^\dagger \xi_2 
\right) =0 \, ,
\]
\[ -\gamma^\mu D_\mu \xi_2 + \frac{2 \pi}{k} \left(
\xi_2 (-1 \frac{k \mu}{2 \pi} +Q_1^\dagger Q_1-Q_2^\dagger Q_2) - 2 \xi_1 Q_2^\dagger Q_1 
+(Q_2 Q_2^\dagger- Q_1 Q_1^\dagger) \xi_2 + 2 Q_1 Q_2^\dagger \xi_1 
\right) =0 \, .
\]
}
Let us write explicitly the equations for $N=2$.
The following notation is used:
\beq \xi_1=\left(\begin{array}{cc}
\xi_{11}  & \xi_{12} \\
\xi_{21}  & \xi_{22} \\
\end{array}\right)  \, , \qquad
 \xi_2= \left(\begin{array}{cc}
\tilde{\xi}_{11}  & \tilde{\xi}_{12} \\
\tilde{\xi}_{21}  & \tilde{\xi}_{22} \\
\end{array}\right)  \, .
\eeq
We get two systems of two coupled equations and four decoupled equations:
\beq -\gamma^\mu \partial_\mu \xi_{11} -i \gamma^0 \frac{f'}{2 r \mu} \xi_{11}-
\mu (\xi_{11} (1-\psi^2)+2 e^{i \varphi} \psi \tilde{\xi}_{12})=0 \, , \label{sy1}\eeq
\[ -\gamma^\mu \partial_\mu \tilde{\xi}_{12} -i (1-f) \frac{x \gamma^2-y \gamma^1}{r^2} \tilde{\xi}_{12} -
2 \mu e^{-i \varphi} \psi \xi_{11}=0 \, .\]
\beq -\gamma^\mu \partial_\mu \tilde{\xi}_{22} +i \gamma^0 \frac{f'}{2 r \mu} \tilde{\xi}_{22}-
\mu (\tilde{\xi}_{22} (1-\psi^2)+2 e^{-i \varphi} \psi \xi_{21})=0 \, , \label{sy2} \eeq
\[ -\gamma^\mu \partial_\mu \xi_{21} +i (1-f) \frac{x \gamma^2-y \gamma^1}{r^2} \xi_{21} -
2 \mu e^{i \varphi} \psi \tilde{\xi}_{22}=0 \, .\]
\beq -\gamma^\mu \partial_\mu \xi_{22}+i \gamma_0 \frac{f'}{2 r \mu} \xi_{22} - \mu (1+\psi^2) \xi_{22} = 0 \, , \label{e1} \eeq
\beq   -\gamma^\mu \partial_\mu \tilde{\xi}_{11} -i \gamma^0 \frac{f'}{2 r \mu} \tilde{\xi}_{11}- \mu (1+\psi^2) \tilde{\xi}_{11}=0 \, ,  \label{e2} \eeq
\beq -\gamma^\mu \partial_\mu \xi_{12} -i (1-f) \frac{x \gamma^2-y \gamma^1}{r^2}  \xi_{12} -2 \mu \xi_{12}= 0 \, , \label{e3} \eeq
\beq -\gamma^\mu \partial_\mu \tilde{\xi}_{21} +i (1-f) \frac{x \gamma^2-y \gamma^1}{r^2} \tilde{\xi}_{21} -2 \mu \tilde{\xi}_{21} = 0 \, \label{e4} .\eeq

It is straightforward to check that Eqs. (\ref{e1}), (\ref{e2}) have no square-integrable solutions.
 Using the BPS equations, we get the system:
\beq -2 \mu \xi_{22}^+ +(-\partial_1 +i \partial_2 )  \xi_{22}^- =0 \, \qquad
(-\partial_1 -i \partial_2 )  \xi_{22}^+ -2 \mu \psi^2  \xi_{22}^-  =0 \, ,
 \eeq
then acting with $(\partial_1 +i \partial_2 ) $ on the first equation 
we get $(\partial_1^2 + \partial_2^2 ) \xi_{22}^- + 4 \mu^2 \psi^2 \xi_{22}^-=0$, which has no
square-integrable solutions.
Eqs.  (\ref{e3}), (\ref{e4}) also does not give any zero modes, they correspond to a 2-dimensional
fermion with a Dirac mass term in the background of a vortex (this case is studied in \cite{deVega}).

Let us go back to the two systems (\ref{sy1}), (\ref{sy2}). 
They are trivially related by a complex
conjugation.
Let us use the variables $\eta=\xi_{11},\tilde{\xi}_{22}^*$ 
and $\lambda=\tilde{\xi}_{12},\xi_{21}^*$:
\[ 
\left(\begin{array}{cc}
-\partial_2  & -\partial_1 \\
-\partial_1  & \partial_2 \\
\end{array}\right)   \left(\begin{array}{c} \eta_+ \\ \eta_- \end{array}\right)
-\frac{f'}{2 r \mu} \left(\begin{array}{cc}
0  & i \\
-i  & 0 \\
\end{array}\right) \left(\begin{array}{c} \eta_+ \\ \eta_- \end{array}\right)
 - \mu(1-\psi^2) \left(\begin{array}{c} \eta_+ \\ \eta_- \end{array}\right)
-2 \mu \psi e^{i \varphi}  \left(\begin{array}{c} \lambda_+ \\ \lambda_- \end{array}\right) =0 \, , 
\]
\beq \left(\begin{array}{cc}
-\partial_2  & -\partial_1 \\
-\partial_1  & \partial_2 \\
\end{array}\right)   \left(\begin{array}{c} \lambda_+ \\ \lambda_- \end{array}\right) 
-i \frac{1-f}{r}   \left(\begin{array}{cc}
x/r  & -y/r \\
-y/r  & -x/r \\
\end{array}\right) \left(\begin{array}{c} \lambda_+ \\ \lambda_- \end{array}\right) 
-2 \mu \psi e^{-i \varphi} \left(\begin{array}{c} \eta_+ \\ \eta_- \end{array}\right) = 0 \, . \label{fermione}
\eeq
After a change of the $\gamma$ matrices basis, the equations become:
{\small \[ 
\left(\begin{array}{cc}
0 & -\partial_1-i \partial_2 \\
-\partial_1 + i \partial_2 & 0\\
\end{array}\right)   \left(\begin{array}{c} \eta_+ \\ \eta_- \end{array}\right)
-\frac{f'}{2 r \mu} \left(\begin{array}{cc}
-1  & 0 \\
0  & 1 \\
\end{array}\right) \left(\begin{array}{c} \eta_+ \\ \eta_- \end{array}\right)
 - \mu(1-\psi^2) \left(\begin{array}{c} \eta_+ \\ \eta_- \end{array}\right)
-2 \mu \psi e^{i \varphi}  \left(\begin{array}{c} \lambda_+ \\ \lambda_- \end{array}\right) =0 \, , 
\]
\beq \left(\begin{array}{cc}
0 & -\partial_1-i \partial_2 \\
-\partial_1 + i \partial_2 & 0\\
\end{array}\right)   \left(\begin{array}{c} \lambda_+ \\ \lambda_- \end{array}\right) 
+ \frac{1-f}{r}   \left(\begin{array}{cc}
0  & e^{i \varphi} \\
-e^{-i \varphi}  & 0 \\
\end{array}\right) \left(\begin{array}{c} \lambda_+ \\ \lambda_- \end{array}\right) 
-2 \mu \psi e^{-i \varphi} \left(\begin{array}{c} \eta_+ \\ \eta_- \end{array}\right) = 0 \, .
\eeq }
The problem is reduced to the one of finding the kernel of the operator:
{\small \beq \mathcal{D}=
\left(\begin{array}{cccc}
-\partial_1+i \partial_2  & -2 \mu \psi e^{i \varphi} & 0 & 0\\
-2 \mu \psi e^{-i \varphi}  & -\partial_1-i \partial_2 +\frac{1-f}{r} e^{i \varphi}  & 0 & 0\\
2 \mu (\psi^2-1 )  & 0  & -\partial_1- i \partial_2&  -2 \mu \psi e^{i \varphi}\\
0 & 0 & -2 \mu \psi e^{-i \varphi} & -(\partial_1-i\partial_2)-\frac{1-f}{r} e^{-i \varphi} \\ 
\end{array}\right) \, ,
\eeq }
acting on $(\eta_+,\lambda_-,\eta_-,\lambda_+)^t$.
Let us introduce the ausiliary operators
\beq \mathcal{D}_1=\left(\begin{array}{cc} 
-\partial_1+i \partial_2  & -2 \mu \psi e^{i \varphi}  \\ 
-2 \mu \psi e^{-i \varphi}  & -\partial_1-i \partial_2 +\frac{1-f}{r} e^{i \varphi}  \\
 \end{array}\right) \, , \eeq
\[ \mathcal{D}_2=\left(\begin{array}{cc} 
-\partial_1- i \partial_2& - 2 \mu \psi e^{i \varphi}\\ 
-2 \mu \psi e^{-i \varphi} & -(\partial_1-i\partial_2)-\frac{1-f}{r} e^{-i \varphi}  \\
 \end{array}\right) \, .\]
 A computation 
with the index theorem \cite{WeinbergA,WeinbergB} tell us that 
\[ {\rm dim }({\rm kernel} \mathcal{D})- {\rm dim }({\rm kernel} \mathcal{D}^\dagger)=0\, . \]
The operators $ \mathcal{D}_1$ and  $ \mathcal{D}_2^\dagger$ have a trivial kernel;
index theorem then can be used to show that $ \mathcal{D}_1^\dagger$ and  $ \mathcal{D}_2$ have a kernel
with real dimension two. This shows that
\[ {\rm dim }({\rm kernel} \mathcal{D})= {\rm dim }({\rm kernel} \mathcal{D}^\dagger)=2 \, .\]
A related calculation can be found in \cite{LeeB}.
We get 4 fermionic zero modes from the $\xi$ sector.

\subsection{$\chi$ sector}
The relevant fermionic action is:
\beq -i \, \Tr (\chi^\dagger)^I \gamma^\mu D_\mu \chi_I     
 +\frac{2 \pi i}{k} \, \Tr \left(
( Q_1^\dagger Q_1  +Q_2^\dagger Q_2  + 1 \frac{k \mu}{2 \pi}  ) (\chi_1^\dagger \chi_1+\chi_2^\dagger \chi_2)-
\right. \eeq
\[ \left. 
- \chi_1^\dagger (Q_1 Q_1^\dagger+Q_2 Q_2^\dagger ) \chi_1 -
 \chi_2^\dagger (Q_1 Q_1^\dagger+Q_2 Q_2^\dagger ) \chi_2 
+Q_1^\dagger \chi_2 Q_2^\dagger \chi_1 - Q_1^\dagger \chi_1 Q_2^\dagger \chi_2 -  \right. 
\]
\[ \left. 
- Q_2^\dagger \chi_2 Q_1^\dagger \chi_1 + Q_2^\dagger \chi_1 Q_1^\dagger \chi_2 
-Q_1 \chi_2^\dagger  Q_2 \chi_1^\dagger + Q_1 \chi_1^\dagger Q_2 \chi_2^\dagger 
+ Q_2 \chi_2^\dagger Q_1 \chi_1^\dagger - Q_2 \chi_1^\dagger Q_1 \chi_2^\dagger 
  \right) \, . \]
The Dirac equations follow:
{ \small
\[ -\gamma^\mu D_\mu \chi_1 + \frac{2 \pi}{k} \left(
\chi_1 (1 \frac{k \mu}{2 \pi} +Q_1^\dagger Q_1+Q_2^\dagger Q_2) 
-(Q_1 Q_1^\dagger+ Q_2 Q_2^\dagger) \chi_1 
+2 Q_2. \chi_2^\dagger  .Q_1 - 2 Q_1. \chi_2^\dagger  .Q_2
\right) =0
\]
\[ -\gamma^\mu D_\mu \chi_2 + \frac{2 \pi}{k} \left(
\chi_2 (1 \frac{k \mu}{2 \pi} +Q_1^\dagger Q_1+Q_2^\dagger Q_2) 
-(Q_1 Q_1^\dagger+ Q_2 Q_2^\dagger) \chi_2 
-2 Q_2. \chi_1^\dagger  .Q_1 + 2 Q_1. \chi_1^\dagger  .Q_2
\right) =0
\]
}
Let us again specialize to $N=2$.
The following notation is used:
\beq \chi_1=\left(\begin{array}{cc}
\chi_{11}  & \chi_{12} \\
\chi_{21}  & \chi_{22} \\
\end{array}\right)  \, , \qquad
 \chi_2= \left(\begin{array}{cc}
\tilde{\chi}_{11}  & \tilde{\chi}_{12} \\
\tilde{\chi}_{21}  & \tilde{\chi}_{22} \\
\end{array}\right)  \, .
\eeq 
The details of the calculations are rather similar to the ones for the $\xi$ sector.
We get two systems of two coupled equations and four decoupled equations:
\beq -\gamma^\mu \partial_\mu \chi_{22} +i \gamma^0 \frac{f'}{2 r \mu} \chi_{22}
+ \mu ( (1-\psi^2)  \chi_{22} +2 \psi e^{i \varphi} \tilde{\chi}_{21}^* ) =0 \, , \label{sy3} \eeq
\[  -\gamma^\mu \partial_\mu \tilde{\chi}_{21} +i (1-f) \frac{x \gamma^2-y \gamma^1}{r^2} \tilde{\chi}_{21} +
2 \mu \psi e^{i \varphi} \chi_{22}^* =0 \, .   \]
\beq  -\gamma^\mu \partial_\mu \tilde{\chi}_{22} +i \gamma^0 \frac{f'}{2 r \mu} \tilde{\chi}_{22}
+ \mu (  (1-\psi^2 ) \tilde{\chi}_{22}  - 2 \psi e^{i \varphi} \chi_{21}^* )=0 \, . \label{sy4} \eeq
\[  -\gamma^\mu \partial_\mu \chi_{21} +i (1-f) \frac{x \gamma^2-y \gamma^1}{r^2} \chi_{21}
-2 \mu \psi e^{i \varphi} \tilde{\chi}_{22}^* =0 \, ,
\]
\beq  -\gamma^\mu \partial_\mu \chi_{11} -i \gamma^0 \frac{f'}{2 r \mu} \chi_{11}
+\mu (1+\psi^2) \chi_{11} =0 \, , \label{e5} \eeq
\beq  -\gamma^\mu \partial_\mu \tilde{\chi}_{11} -i \gamma^0 \frac{f'}{2 r \mu} \tilde{\chi}_{11}
+\mu (1+\psi^2) \tilde{\chi}_{11} =0 \, , \label{e6} \eeq
\beq -\gamma^\mu \partial_\mu \chi_{12} -i (1-f) \frac{x \gamma^2-y \gamma^1}{r^2}  \chi_{12} +2 \mu \chi_{12}= 0 \, ,  \label{e7}  \eeq
\beq -\gamma^\mu \partial_\mu \tilde{\chi}_{12} -i (1-f) \frac{x \gamma^2-y \gamma^1}{r^2}  \tilde{\chi}_{12} +2 \mu \tilde{\chi}_{12}= 0 \, .  \label{e8}\eeq
The four decoupled equations (\ref{e5})-(\ref{e8}) have no square-integrable solutions.

The two systems (\ref{sy3}) and (\ref{sy4}) are equivalent.
Let us use the variables $\eta=\chi_{22},\tilde{\chi}_{22}$ and
 $\lambda=\tilde{\chi}_{21}^*,\chi_{21}^*$:
\[ 
\left(\begin{array}{cc}
-\partial_2  & -\partial_1 \\
-\partial_1  & \partial_2 \\
\end{array}\right)   \left(\begin{array}{c} \eta_+ \\ \eta_- \end{array}\right)
+\frac{f'}{2 r \mu} \left(\begin{array}{cc}
0  & i \\
-i  & 0 \\
\end{array}\right) \left(\begin{array}{c} \eta_+ \\ \eta_- \end{array}\right)
 + \mu(1-\psi^2) \left(\begin{array}{c} \eta_+ \\ \eta_- \end{array}\right)
\pm 2 \mu \psi e^{i \varphi}  \left(\begin{array}{c} \lambda_+ \\ \lambda_- \end{array}\right) =0 \, , 
\]
\beq \left(\begin{array}{cc}
-\partial_2  & -\partial_1 \\
-\partial_1  & \partial_2 \\
\end{array}\right)   \left(\begin{array}{c} \lambda_+ \\ \lambda_- \end{array}\right) 
-i \frac{1-f}{r}   \left(\begin{array}{cc}
x/r  & -y/r \\
-y/r  & -x/r \\
\end{array}\right) \left(\begin{array}{c} \lambda_+ \\ \lambda_- \end{array}\right) 
\pm 2 \mu \psi e^{-i \varphi} \left(\begin{array}{c} \eta_+ \\ \eta_- \end{array}\right) = 0 \, .
\eeq
Changing the $\gamma$ matrices basis:
{\small \[ 
\left(\begin{array}{cc}
0 & -\partial_1-i \partial_2 \\
-\partial_1 + i \partial_2 & 0\\
\end{array}\right)   \left(\begin{array}{c} \eta_+ \\ \eta_- \end{array}\right)
-\frac{f'}{2 r \mu} \left(\begin{array}{cc}
1  & 0 \\
0  & -1 \\
\end{array}\right) \left(\begin{array}{c} \eta_+ \\ \eta_- \end{array}\right)
 + \mu(1-\psi^2) \left(\begin{array}{c} \eta_+ \\ \eta_- \end{array}\right)
\pm 2 \mu \psi e^{i \varphi}  \left(\begin{array}{c} \lambda_+ \\ \lambda_- \end{array}\right) =0 \, , 
\]
\beq \left(\begin{array}{cc}
0 & -\partial_1-i \partial_2 \\
-\partial_1 + i \partial_2 & 0\\
\end{array}\right)   \left(\begin{array}{c} \lambda_+ \\ \lambda_- \end{array}\right) 
+ \frac{1-f}{r}   \left(\begin{array}{cc}
0  & e^{i \varphi} \\
-e^{-i \varphi}  & 0 \\
\end{array}\right) \left(\begin{array}{c} \lambda_+ \\ \lambda_- \end{array}\right) 
\pm 2 \mu \psi e^{-i \varphi} \left(\begin{array}{c} \eta_+ \\ \eta_- \end{array}\right) = 0 \, .
\eeq }
The problem is reduced to the one of finding the kernel of the following operator:
{\small \beq \mathcal{D}=
\left(\begin{array}{cccc}
-\partial_1+i \partial_2  & \pm 2 \mu \psi e^{i \varphi} & 0 & 0\\
\pm 2 \mu \psi e^{-i \varphi}  & -\partial_1-i \partial_2 +\frac{1-f}{r} e^{i \varphi}  & 0 & 0\\
2 \mu (1-\psi^2 )  & 0  & -\partial_1- i \partial_2&  \pm 2 \mu \psi e^{i \varphi}\\
0 & 0 & \pm 2 \mu \psi e^{-i \varphi} & -(\partial_1-i\partial_2)-\frac{1-f}{r} e^{-i \varphi} \\ 
\end{array}\right) \, .
\eeq }
The following ausiliary operators are introduced:
\beq \mathcal{D}_1=\left(\begin{array}{cc} 
-\partial_1+i \partial_2  & \pm 2 \mu \psi e^{i \varphi}   \\ 
\pm 2 \mu \psi e^{-i \varphi}  & -\partial_1-i \partial_2 +\frac{1-f}{r} e^{i \varphi}  \\
 \end{array}\right) \, .\eeq
\[ \mathcal{D}_2=\left(\begin{array}{cc} 
-\partial_1- i \partial_2&  \pm 2 \mu \psi e^{i \varphi}\\ 
 \pm 2 \mu \psi e^{-i \varphi} & -(\partial_1-i\partial_2)-\frac{1-f}{r} e^{-i \varphi}  \\
 \end{array}\right) \, .\]
 A computation 
with the index theorem also tell us that 
\[{\rm dim }({\rm kernel} \mathcal{D})- {\rm dim }({\rm kernel} \mathcal{D}^\dagger)=0 \, . \]
The operators $ \mathcal{D}_1$ and  $ \mathcal{D}_2^\dagger$ have a trivial kernel;
index theorem then can be used to show that $ \mathcal{D}_1^\dagger$ and  $ \mathcal{D}_2$ have a kernel
with real dimension two. We get a total of four zero modes 
from the $\chi$ sector.

\end{document}